\documentclass[sigconf,screen]{acmart}

\AtBeginDocument{%
  \providecommand\BibTeX{{%
    \normalfont B\kern-0.5em{\scshape i\kern-0.25em b}\kern-0.8em\TeX}}}

\setcopyright{acmcopyright}
\copyrightyear{2018}
\acmYear{2018}
\acmDOI{XXXXXXX.XXXXXXX}

\acmConference[Conference acronym 'XX]{Make sure to enter the correct
  conference title from your rights confirmation emai}{June 03--05,
  2018}{Woodstock, NY}

\acmBooktitle{Woodstock '18: ACM Symposium on Neural Gaze Detection,
 June 03--05, 2018, Woodstock, NY} 
\acmPrice{15.00}
\acmISBN{978-1-4503-XXXX-X/18/06}

\usepackage{array}
\usepackage{multirow}
\usepackage{subcaption}
\usepackage{graphicx}
\usepackage[export]{adjustbox}
\usepackage{amsmath}

\usepackage[normalem]{ulem}
\usepackage{threeparttable}
\usepackage[table,xcdraw]{xcolor}  
\usepackage{colortbl} 

\usepackage{xcolor, soul}
\renewcommand\hl[1]{#1} 

\definecolor{myorange}{rgb}{1, 0.647, 0}  

\newcommand{\hlorange}[1]{\begingroup\sethlcolor{myorange}\hl{#1}\endgroup}

\begin{document}

\title[Dream the Dream]{Dream the Dream: Futuring Communication between LGBTQ+ and Cisgender Groups in Metaverse}


\author{Anqi Wang}
\affiliation{%
  \institution{Hong Kong University of Science and Technology}
  \country{Hong Kong SAR}
}

\author{Lei Han}
\affiliation{%
  \institution{Hong Kong University of Science and Technology (Guangzhou)}
  \city{Guangzhou}
  \country{China}
}

\author{Jiahua Dong}
\affiliation{%
  \institution{The Chinese University of Hong Kong}
  \country{Hong Kong SAR}
}

\author{Muzhi Zhou}
\affiliation{%
  \institution{Hong Kong University of Science and Technology (Guangzhou)}
  \city{Guangzhou}
  \country{China}
}

\author{David Yip}
\affiliation{%
  \institution{Hong Kong University of Science and Technology (Guangzhou)}
  \city{Guangzhou}
  \country{China}
}

\author{Yuyang Wang}
\affiliation{%
  \institution{Hong Kong University of Science and Technology (Guangzhou)}
  \city{Guangzhou}
  \country{China}
}

\author{Pan Hui}
\authornote{Corresponding author.}
\affiliation{%
  \institution{Hong Kong University of Science and Technology (Guangzhou)}
  \city{Guangzhou}
  \country{China}
}
\affiliation{%
  \institution{Hong Kong University of Science and Technology}
  \country{Hong Kong SAR}
}

\renewcommand{\shortauthors}{Anqi Wang et al.}

\begin{abstract}
    Digital platforms frequently reproduce heteronormative norms and structural biases, limiting inclusive communication between LGBTQ+ and cisgender individuals. The Metaverse, with its affordances for identity fluidity, presence, and community governance, offers a promising site for reimagining such interactions. To investigate this potential, we conducted participatory design workshops involving LGBTQ+ and cisgender participants, situating them in speculative Metaverse contexts to surface barriers and co-create alternative futures. The workshops followed a three-phase process—identifying challenges, speculative problem-solving, and visualizing futures—yielding socio-spatial-technical solutions across four layers: activity, interaction, scene, and space. These findings highlight the importance of spatial cues and power dynamics in shaping digital encounters. We contribute by (1) articulating challenges of cross-group communication in virtual environments, (2) proposing inclusive design opportunities for the Metaverse, and (3) advancing principles for addressing power geometry in digital space. This work demonstrates futuring as a critical strategy for designing equitable, transformative communication infrastructures. 
\end{abstract}




\begin{teaserfigure}
  \includegraphics[width=\textwidth]{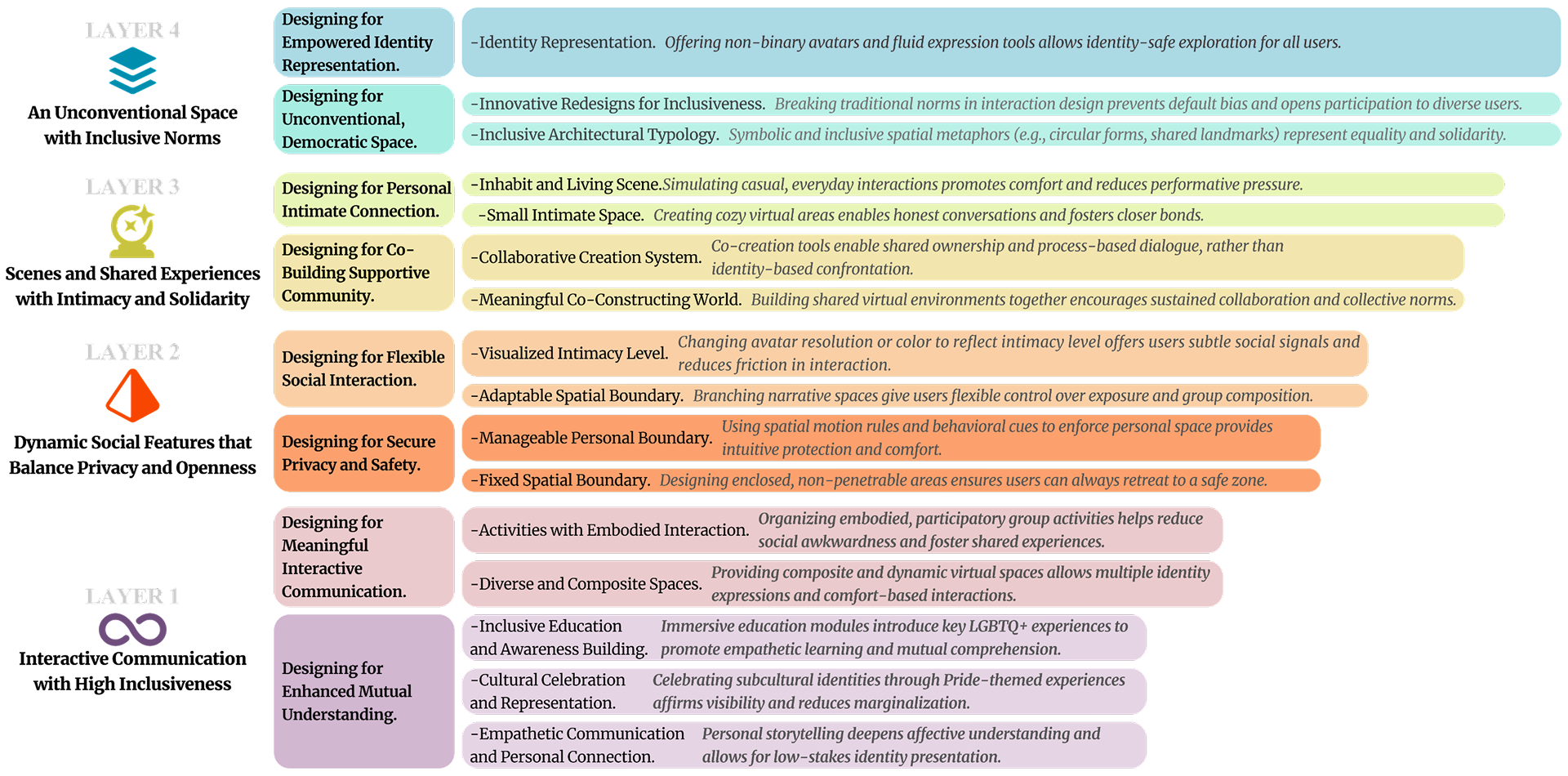}
  \caption{Speculating future Metaverse that bridge the communication between LGBTQ+ and cisgender groups.}
  \Description{Speculating future Metaverse that bridge the communication between LGBTQ+ and cisgender groups.}
  \label{fig:teaser}
\end{teaserfigure}


\maketitle

\section{Introduction}
    The Metaverse, encompassing evolving concepts such as video games, visual design, and immersive environments~\mbox{\cite{10035386}}, offers new opportunities for marginalized people~\mbox{\cite{mao2024review}}, such as LGBTQ+ people, by critiquing current technologies speculation and design. 
LGBTQ+ refers to lesbian, gay, bisexual, transgender, and queer, with the ``+'' signifying the complexity and open-ended development of sexual and gender minorities~\mbox{\cite{parent2013approaches}}. 
While social technologies provide boundary-less communication channels, such as supporting identity expression~\mbox{\cite{transonyoutube,cavalcante2017did}} and community building~\mbox{\cite{jenzen2014make}}, research shows that digital platforms often issues, such as misunderstandings due to single modalities~\mbox{\cite{Vivienne01072012}}, over-censored~\mbox{\cite{monea2022digital,duffy2021nested}} and silenced~\mbox{\cite{10.1145/3479577,10.1145/3555209,10.1145/3555149,10.1145/3555105}}. This all reflects reproduced heteronormative designs and reinforced unequal power dynamics from offline social norms~\mbox{\cite{scheuerman2018safe,walker2020more,gray2014race,mckenna2020resistance}}, leading to persistent misinformation, trolling, and the reinforcement of harmful stereotypes~\mbox{\cite{scheuerman2018safe,monea2022digital,wang2022gay,10.1145/3491102.3517624}}.

\hl{While prior work has explored how virtual spaces support LGBTQ+ communities internally~\mbox{\cite{hardy2022lgbtqfuture,hardy2019participatory,haimson2020designing}}, we lack understanding of \textbf{what design features could facilitate communication between LGBTQ+ and cisgender people}, particularly in where heteronormative dominance creates significant barriers.}
While social psychology research—most notably \textit{Contact Theory}~\cite{allport1954nature}—emphasizes the importance of structured and equitable interaction conditions for prejudice reduction, 
mainstream digital platforms seldom provide such conditions. Instead, platform affordances such as anonymity, weak accountability mechanisms, and algorithmic governance frequently entrench asymmetric visibility and power relations, particularly for marginalized users~\cite{10.1145/3555105,10.1145/3479577}.
~
The Metaverse thus presents a promising but underexplored design space. While its affordances enable new forms of presence and social ordering, there remains a lack of design-oriented accounts that articulate \textit{how socio-spatial affordances can be composed to mediate power relations and support inclusive cross-group communication.}

To address this challenge, we employ \emph{Future Design} method in workshops that involve LGBTQ+ and cisgender individuals in envisioning communication futures within Metaverse spaces~\mbox{\cite{bossen2016evaluation}}. 
Grounded in prior HCI research~\mbox{\cite{hardy2022lgbtqfuture,haimson2020designing,cheon_i_2019,zheng2024charting,harrington2021eliciting,10.1145/3613904.3642609}}, \textit{Future Design} methods explore speculative and participatory approaches to alternative futures~\mbox{\cite{lindley2015back}}. Recent work has also examined the Metaverse as future-oriented lenses to challenge bias and support marginalized experiences using speculation~\mbox{\cite{10.1145/3613904.3641972,10.1145/3544548.3581306,10.1145/3643834.3661637,10.1145/3686939,10.1145/3544548.3581305,hardy2022lgbtqfuture,10.1145/3301019.3323894,solyst2023would,harrington2021eliciting,haimson2020designing}}. 
\hl{``Futuring'' within the Metaverse offers a means of addressing such challenges by envisioning alternative social orders and reconfiguring inclusion and exclusion, foregrounding fluid identity representation, embodied presence, and self-governed communities~\mbox{\cite{10.1145/3688569,10.1007/s43762-022-00051-0,10302997}}.} 
\hl{
This democratic design approach promotes mutual learning and empowers LGBTQ+ participation, addressing the identified gap by focusing on the diverse needs of different populations.}


~
To promote mutual understanding and foreground LGBTQ+ voices, the workshops followed three phases: 
\textit{(1) Identifying Challenges}, 
\textit{(2) Speculative Problem-Solving}, and 
\textit{(3) Visualizing Futures}. 
This process generated socio-spatial-technical solutions across four layers: (1) Activity—interactive and educational communication strategies; (2) Interaction—social features affording privacy and intimacy; (3) Scene—future scenarios fostering solidarity; and (4) Space—unconventional environments grounded in inclusive norms (Figure~\ref{fig:teaser}). 
These findings highlight the role of spatial cues in shaping interactions and social proximity (Section~\ref{sec:designsuggestions}) and are further synthesized through the framework of power geometry, which examines how unequal power dynamics structure virtual spaces (Section~\ref{sec:powergeometry}). 
~
\hl{This paper makes three contributions to DIS and HCI: 
(1) \textbf{Empirical Contribution:} We reframe cross-group communication as a socio-spatial design problem, shifting attention from interfaces to how space, presence, and interaction structures mediate power relations.
(2) \textbf{Design Contribution:} Through joint-group future workshops, we articulate socio-spatial design strategies—specifically dynamic proxemics and collaborative co-presence——that address communication barriers across spatial, interactional, and relational dimensions in future Metaverse environments.
(3) \textbf{Theoretical Contribution:} We propose transferable design principles that position futuring as an infrastructural strategy for designing Metaverse spaces attentive to power asymmetries and inclusion.} 



\section{Related Works}
    


To situate our work, we focus on \textit{virtual space} rather than the speculative \textit{Metaverse}. Immersive environments—video games, multiplayer VR, social VR, and virtual worlds—serve as early prototypes of the Metaverse. Examining how LGBTQ+ individuals navigate these spaces, face communication challenges, and build communities provides empirical insight into potential LGBTQ+ communication opportunities in future Metaverse environments. 
This section introduces the Metaverse as a radical future concept and reviews current virtual spaces to identify communication opportunities between LGBTQ+ and cisgender participants.

\subsection{Metaverse, a Speculative Future}~\label{sec:Metaversedefinition}
\hl{
The \textit{Metaverse} is an unrealized vision extending beyond current technology, society, and human experience. The prefix ``meta'' (transcendence) and suffix ``verse'' (dimension of time and space) denote a synthetic environment linking physical and virtual worlds. Early frameworks highlight identity presence, ubiquity of access, interoperability, and scalability~\mbox{\cite{10.1145/2480741.2480751}}. More recent accounts depict it as the next-generation Internet, where users interact as avatars within convergent technological ecosystems~\mbox{\cite{10.1007/s43762-022-00051-0}}. 

Roadmaps emphasize technical requirements including scalability, accessibility, security, privacy, and legal frameworks~\mbox{\cite{9877927}}, positioning the Metaverse as a potential catalyst for social transformation~\mbox{\cite{smartcities5030043}}. Practitioners contribute utopian explorations of nation-states, corporations, virtual life forms, property, and monetized economies~\mbox{\cite{Sönmez2023}}. Scholars note its capacity to include, exclude, establish, and disrupt social orders~\mbox{\cite{10.1145/3688569,10.1007/s43762-022-00051-0,10302997}}. 

Central to these discussions is the ``humanization'' of virtual environments. Zallio et al. identify three pillars—tangible realities (e.g., companies, nation-states, assets), worldviews (e.g., liberalism, socialism, religion), and emotional states (e.g., happiness, freedom, safety, love, sex)—highlighting its influence on societal structures and personal experience~\mbox{\cite{ZALLIO2022101909}}. The Metaverse is envisioned as inclusive, diverse, accessible, safe, and equitable, with success dependent not only on technology but also on empathetic design principles~\mbox{\cite{ZALLIO2022101909}}. Compared to VR-based virtual spaces, it represents a high-tech ecosystem with broader societal impact.

}

\subsection{Virtual Space as a Solution to Communication Challenges}
\textit{Virtual space} provides unique features, including embodied avatars, simulated interactions, rich content, and customizable environments~\cite{zheng2023understanding}. Avatars enable users to express individual characteristics beyond gender binaries~\mbox{\cite{baker2021avatar,acena2021my,freeman2021body,freeman2022acting}}, while immersive content and flexible environments enhance engagement and intimacy~\mbox{\cite{divine2020falling,li2023wecried,joshua2018ecology}}. These spaces serve as havens for marginalized groups, facilitating connections and overcoming identity visibility and safety challenges~\cite{cabiria2008virtual,doring2009internet,campkin2017lgbtq+}.

\subsubsection{Safety Design}
Safety designs in virtual spaces are crucial for fostering trust and addressing communication challenges. Techniques such as boundary settings, rapid responses, and agreements have proven effective~\cite{zheng2023understanding}. Many designs are grounded in proxemics theory, which posits that individuals maintain personal space to ensure comfort and reduce anxiety~\cite{edward1968proxemics}. For example, \textit{VRChat} allows users to customize settings based on trust rankings, controlling voice, avatar elements, interface tools, and visual effects~\cite{VRChatSafeTrustSystem}. Similarly, \textit{QuiVR} employs a ``personal bubble'' to fade avatars who intrude and a ``power gesture'' to remove nearby players~\cite{McAloon2016QuiVR}, demonstrating proxemics-informed safety.

Smaller virtual spaces further enhance safety by supporting cohesive group formation, shared attention, and personal space preservation~\cite{edward1968proxemics}. Nonetheless, safety challenges often stem from power dynamics in interactions between LGBTQ+ and cisgender users~\cite{scheuerman2018safe,oreglia2016ict,mckenna2020resistance} and within LGBTQ+ communities~\cite{walker2020more}. Current approaches focusing on ``bad actors'' or safe spaces are limited~\cite{scheuerman2018safe,zheng2023understanding,walker2020more}. Scheuerman et al.~\cite{scheuerman2018safe} advocate for leveraging virtual spaces' structural potential to address embedded power relations, promoting inclusive and equitable interactions through design and governance.

\subsubsection{Identity Management}
Virtual spaces enable flexible identity representations, enhancing visibility and communication across social groups. Customizable avatars—including gender, race, hairstyle, and accessories~\cite{freeman2022re}—allow LGBTQ+ users to express identities more fully than in physical spaces~\cite{pullen2010lgbt}. Embodied identity visibility fosters self-reflection and psychological well-being~\cite{freeman2022re, schulenberg2023creepy, 10.1145/3025453.3025602} and can reduce intergroup prejudice under favorable conditions~\cite{allport1954nature, mereish2015effects, doi:10.1080/14675986.2022.2080969}.

A key strategy is embodied selective visibility, where users choose whether to disclose sexual orientation or identity, emphasizing physical enactment through avatars to experience agency~\cite{woodland2000queer, freeman2022acting}. Rich identity attributes and fluid meanings, common in games and virtual worlds~\cite{kafai2016diversifying, 10.1145/2658537.2658685}, allow safe identity exploration~\cite{acena2021my}. Yet, avatar systems often limit non-binary and transgender expression, failing to capture evolving appearances, legal statuses, and societal perceptions~\cite{freeman2022acting, walker2020more, haimson2020designing}. Adaptable voice features partially address this, but comprehensive infrastructures are needed. Projects like \textit{Virtual Drag} highlight dynamic queer performativity and fluid embodiment~\cite{doi:10.1080/14675986.2022.2080969}, yet no system fully supports broad LGBTQ+ identity variability.

\subsubsection{Activities and Engagements}
Social interactions and activities help address communication challenges, consistent with Allport's contact hypothesis~\cite{allport1954nature}. Strategies include self-presentation, diverse gatherings~\cite{mckenna2020resistance}, user autonomy~\cite{OculusQuest2021lgbtq}, and collaborative efforts~\cite{freeman2022working,OculusQuest2021lgbtq}, fostering dynamic relationships~\cite{ducheneaut2007life}. LGBTQ+ communities organize around identity development, information exchange, and social support, prioritizing solidarity and empowerment~\cite{freeman2022acting,li2023wecried,freeman2021body,acena2021my,scheuerman2018safe,mckenna2020resistance,ducheneaut2007life,doi:10.1177/146144804047513}.

Virtual worlds host numerous social movements and celebratory events, e.g., Pride Month activities in \textit{Mozilla Hubs}, \textit{VRChat}, \textit{Horizon}, and \textit{AltspaceVR}, as well as drag shows in \textit{Horizon}~\cite{OculusQuest2021lgbtq}. Collaborative VR activities support connections often unattainable offline, addressing power inequalities~\cite{mckenna2020resistance}, while real-time verbal and nuanced non-verbal communication (hugging, working) enhances embodied interaction~\cite{freeman2021hugging,freeman2022working,divine2020falling,joshua2018ecology,fare2018smallgroup}. Key features of social VR-mediated collaboration include embodiment, replication of offline activities, and integration of work, play, and daily life~\cite{freeman2022working}.

Educational VR activities further offer immersive, accessible, and perspective-shifting experiences. Narrative-based virtual spaces provide queer education~\cite{geer2023sea, hardy2019participatory}, fostering empathy and prosocial behavior~\cite{doi:10.1080/15213269.2012.755877,FAUVILLE202091,doi:10.1080/15534510802643750}. VR exhibitions, such as interactive museums highlighting LGBTQ+ history~\cite{leslie2022vr}, enhance awareness, though they may oversimplify complex real-world interactions and fail to capture critical emotional and social nuances~\cite{nakamura2020feeling}.

\subsection{Virtual Space as an Amplifier for Communication Challenges} 
\subsubsection{Social Prejudice and Discrimination}
Systemic prejudice amplifies communication challenges between LGBTQ+ and cisgender users~\cite{10.1145/3313831.3376497, scheuerman2018safe}, reinforcing existing power dynamics. Identity disclosure increases vulnerability, as LGBTQ+ users are frequent targets of harassment~\cite{hardy2022lbtqfuture,freeman2022re}. In China, the ``Triple No Policy''—no support, disapproval, or promotion—alongside censorship creates a precarious virtual environment, suppressing open communication and marginalizing LGBTQ+ individuals~\cite{doi:10.1080/01292986.2017.1324500,10.1145/3491102.3517624,2016survey}. A 2021 survey found only 11\% of respondents had publicly outed LGBTQ+ contacts in their networks, far below the global average of 42\%~\cite{2021survey}.


\subsubsection{Heteronormativity Design} 
Heteronormative design, embedding gender binaries into software architecture, further constrains communication~\cite{powell2004data,balsam_measuring_2011,amory2022introduction}. Two primary drivers are limited LGBTQ+ visibility and exclusion from VR design processes. Only 2–3\% of tech professionals identify as LGBTQ+, leading to systemic design biases~\cite{lgbtqdata,li2023wecried}. Algorithmic enforcement, such as social VR trust ranks, may undervalue LGBTQ+ users interacting within their own communities, perpetuating stereotypes and discouraging intergroup communication~\cite{chen_people_2024}. VR security systems often neglect LGBTQ+-specific challenges in handling complaints~\cite{zheng2023understanding}.

Identity management constraints prevent LGBTQ+ users from forming communities or connecting with peers~\cite{carrasco2018queer,podmore2006gone,brubaker2016visibility}. For instance, \textit{Weibo} restricts LGBTQ+ identity validation in institutional verification~\cite{ZHENG2022107021,cui2022we}. Limited gender options, pronoun misuse, and restricted avatar expressions—eye contact, posture—further marginalize users~\cite{freeman2022re}. LGBTQ+-exclusive VR spaces often focus on sexual partner seeking, and safety designs prioritize basic harm prevention over nuanced boundary management~\cite{panfil2019nobody}. Collectively, heteronormative constraints and systemic design deficiencies hinder safe, inclusive virtual environments.

\subsubsection{Virtual Abuse} 
Anonymity in social VR can embolden homophobic behavior, with individuals performing offensive gestures, movements, or speech, inducing discomfort and panic~\cite{walker2020more}. Avatar appearance and voice modes may trigger targeted harassment based on aesthetic preferences~\cite{zheng2023understanding}. Cisgender users can restrict LGBTQ+ participation in events or maliciously crowd them out, for example via vote-kick features~\cite{walker2020more,zheng2023understanding}. LGBTQ+ users also report harassment through impolite curiosity, such as invasive sex-related questions in virtual chats~\cite{woodland2000queer}. In response, gender-camouflaging and androgynous avatars, while protective, can reduce visibility and reinforce marginalization~\cite{schulenberg2023creepy}. 

\subsection{Participatory Design and Design Method for Future Scenarios}
Participatory design involves diverse stakeholders and provides frameworks for actively engaging individuals throughout the design process~\cite{anuyah2023characterizing}. Using a democratic approach, it enables collaborative examination of established knowledge models and power dynamics~\cite{anuyah2023characterizing,Bratteteig2012PD}. This method effectively identifies the critical needs of marginalized populations and reduces subjective design assumptions through collaborative activities such as group brainstorming, questionnaires, and individual interviews~\cite{bossen2016evaluation}. 
With growing emphasis on inclusive social environments, participatory design has been applied to engage vulnerable groups~\cite{comerio1987design, ertner2010five, dindler2020computational, almqvist2023different}, including older adults~\cite{abeele2021immersive, xu2023designing}, Black communities~\cite{solyst_i_2023}, minors~\cite{solyst_i_2023}, and individuals with disabilities~\cite{rui2022online}. However, collaboration with underserved communities requires critical reflection on power and privilege~\cite{gautam_participatory_2018, lazar_making_2018,sabiescu_emerging_2014}, particularly when considering bilateral dynamics involving cisgender participants~\cite{10.5555/1466607.1466609}.

Future-oriented design methods leverage imaginative freedom'' as an alternative to traditional processes to support engagement with marginalized communities~\cite{10.1145/3083671.3083700, 10.1145/2662155.2662194, 10.1145/1900441.1900461}. These methods aim to translate utopian ideas into practical strategies by addressing complexities in future-oriented thinking through empathetic engagement and technology-focused conceptualization~\cite{harrington2021eliciting}. Future workshops, for example, facilitate brainstorming and critique of current applications while including marginalized participants in envisioning potential solutions~\cite{brandt2012tools}. Kensing~\cite{kensing1987generation, kensing_generating_1991} explored participatory approaches framed around what if things could be different'' to amplify marginalized voices and generate feasible solutions. 
In technology-focused contexts, Haimson~\cite{haimson2020designing} conducted collaborative sketching workshops with transgender participants to examine how future technologies could serve resource-poor and low-visibility communities. Similarly, Hardy et al.~\cite{10.1145/3301019.3323894} ran future-oriented participatory design workshops with rural LGBTQ+ groups, combining critique of the present with envisioning and implementation for the future.
Unlike prior studies that focus primarily on LGBTQ+ communities' own experiences, our approach emphasizes future design visions that address communication challenges between LGBTQ+ and cisgender groups. By bringing these communities together, we aim to explore strategies in the Metaverse to enhance mutual understanding, improve accessibility for LGBTQ+ participants, and support specific communication needs. 

\section{Methods}




\hl{
Our study seeks to reduce communication barriers and costs between LGBTQ+ and cisgender people by envisioning a future Metaverse that prioritizes the inclusion of LGBTQ+ voices and suggestions. 
Participatory design has been demonstrated to be an effective approach for fostering collaboration and empathy among diverse stakeholder groups~\mbox{\cite{Bratteteig2012PD}}, as it facilitates mutual understanding and respect among participants.
To address these communication challenges and identify potential solutions for the Metaverse, we conducted three workshops involving both LGBTQ+ and cisgender participants. These workshops employed design activities aimed at creating an inclusive Metaverse, ensuring that the resulting design solutions encapsulate the shared experiences, concerns, and needs of all participants. }

Throughout the workshops, particular attention was given to the marginalized status and individual vulnerabilities of LGBTQ+ participants. This process involved acknowledging conflicting viewpoints, identifying root causes, and examining the power dynamics that influence opinions across different groups. 
\hl{This study received approval from the University Institutional Review Board (IRB). }

    \subsection{Future Workshop with Design Fictions}
Our participatory design process adopted explorative future workshops as the primary methodology. The term ``explorative'' refers to integrating problem identification and problem-solving within the design framework~\cite{schon2017reflective}. This approach requires making strategic decisions about which problems to address. The explorative future design method investigates alternative worlds and future technologies to envision ideal social contexts~\cite{lindley2015back}, aiming to translate these radical visions into actionable design implications.

Given the heteronormative biases inherent in current digital systems, our approach begins with the premise of asking \textit{``what is not on the tech agenda or which solutions are not discussed}~\cite{Bratteteig2012PD}. By \textit{``adding missing elements, we widen the scope and expand the universe of discourse''}~\cite{aas19795}, we develop a vision for our participatory design. Another critical perspective is that mutual learning within participatory design can influence specific power dynamics~\cite{Bratteteig2012PD}, forming the foundation for subsequent problem sets, envisioning exercises, and sketching activities. Mutual learning fosters trust and respect for diverse perspectives~\cite{ehn2014making}, particularly by valuing the expertise and lived experiences of LGBTQ+ people. 
Our study adheres to the guidelines set forth by Bratteteig et al., which emphasize three core values of participatory design: having a say, mutual learning, and co-realization~\cite{Bratteteig2012PD}. 

    \subsection{Recruitment and Participants}
\begin{table*}[ht]
    \small
    \centering
    \begin{threeparttable}
        \caption{Summary of participants' demographics, as reported in the screener.}
        \Description{Summary of participants' demographics, as reported in the screener.}
        \label{tab:participants}
        \begin{tabular*}{\linewidth}{@{\extracolsep{\fill}}p{1.5cm}p{2cm}p{1.8cm}p{2.2cm}p{1.5cm}ccc@{}}
            \toprule
            \textbf{Participants} & \textbf{Workshop} & \textbf{Gender} & \textbf{Sexual Orientation} & \textbf{Age} & \textbf{VR Experience\tnote{1}} \\
            \midrule
            \multirow{6}{*}{\textbf{Workshop 1}} & P1 & Cis Female & Heterosexual & 18-25 & 3 \\
            & P2 & Cis Male & Heterosexual & 18-25 & 2 \\
            & P3 & Cis Female & Lesbian & 18-25 & 2 \\
            & P4 & Cis Male & Gay & 25-40 & 5 \\
            & P5 & Cis Male & Heterosexual & 25-40 & 5 \\
            & P6 & Cis Female & Heterosexual & 18-25 & 1 \\
            \midrule
            \multirow{6}{*}{\textbf{Workshop 2}} & P7 & Cis Male & Gay & 18-25 & 4 \\
            & P8 & Trans Female & Lesbian & 25-40 & 2 \\
            & P9 & Non-Binary & Pansexual & 18-25 & 2 \\
            & P10 & Cis Female & Heterosexual & 18-25 & 4 \\
            & P11 & Cis Male & Heterosexual & 18-25 & 2 \\
            & P12 & Cis Male & Heterosexual & 18-25 & 1 \\
            \midrule
            \multirow{6}{*}{\textbf{Workshop 3}} & P13 & Cis Female & Heterosexual & 18-25 & 3 \\
            & P14 & Cis Female & Bisexual & 18-40 & 2 \\
            & P15 & Cis Female & Lesbian & 18-25 & 2 \\
            & P16 & Cis Male & Heterosexual & 25-40 & 5 \\
            & P17 & Cis Female & Pansexual & 18-25 & 3 \\
            & P18 & Cis Male & Unsure & 25-40 & 2 \\
            \bottomrule
        \end{tabular*}
\begin{tablenotes}
    \small
      \item[1]VR experience is represented on a scale of 5 to 1: Professional, Skilled, Average experience, Minimal experience, and No experience. 
\end{tablenotes}

    \end{threeparttable}
\end{table*}
\subsubsection{Recruitment}
\hl{We recruited participants both online and offline related communities.} 
Recruitment materials included information about the research goal and participatory design workshops approach, emphasizing discussions between LGBTQ+ and cisgender people.
To broaden participation, we encouraged the dissemination of advertisements through various online channels. Simultaneously, offline advertisements were posted on campus bulletin boards and in residential areas. 

An online screening survey accompanied the recruitment advertisement to collect demographic information and determine participants' availability for an in-person workshop. 
\hl{To ensure adequate representation and diversity among participants,}
\hlorange{we included a self-reported scale in this pre-study survey (detail in Table\mbox{\ref{tab:participants}}),
where participants were asked to rate their experience on a 1-5 scale, where: 1 = No experience, 2 = Minimal experience, 3 = Average experience, 4 = Skilled, and 5 = Professional.}
\hl{
Our screening includes those with minimal exposure, as participatory design aims to address the needs of underserved populations~\mbox{\cite{Bratteteig2012PD,gautam_participatory_2018}}. }
\hlorange{Diverse participants would contribute different aspects, ranging from technical aspects of VR to fresh perspectives on accessibility and user-friendly design, based on their level of VR experience.}
Before the formal study, all participants were informed about the study's details and signed a consent form emphasizing their right to withdraw at any time. Confidentiality was also emphasized, with participants assured that numerical codes would anonymize their identities in the study's transcriptions and tangible materials. 

\subsubsection{Participants}
\hl{Table~\mbox{\ref{tab:participants}} reported the 18 participants. They represented diverse gender identities and sexual orientations. The sample size for this qualitative study was determined by reaching theoretical saturation~\mbox{\cite{guest2011applied}}; no new insights emerged from the last two workshops, indicating sufficient data to meet our research objectives. }
Participants recruited in the workshops were aged between 22 and 28, with an average age of 25. The genders identified were as follows: cis-female (8), cis-male (8), \hl{non-binary (1), and transgender (1)}. Regarding sexual orientation, participants identified as heterosexual (9), lesbian (3), gay (2), pansexual (2), bisexual (1), and unsure (1) (Table~\ref{tab:participants}). 
\hl{Given the future workshop's objective of envisioning a fictional, radical future with anti-traditional narratives~\mbox{\cite{10.1145/3083671.3083700,10.1145/1900441.1900461,10.1145/2662155.2662194}}, participants were encouraged—but not required—to have some prior with VR experience. 
All participants would have a 10-minute VR trial before the workshop.}
Each workshop lasted approximately 120 minutes and included six participants, with at least two LGBTQ+ participants representing differing sexual orientations.

    \subsection{Participatory Design Workshops}
\begin{figure*}[ht]
    \begin{subfigure}{0.45\textwidth}
        \centering
        \includegraphics [height=5cm]{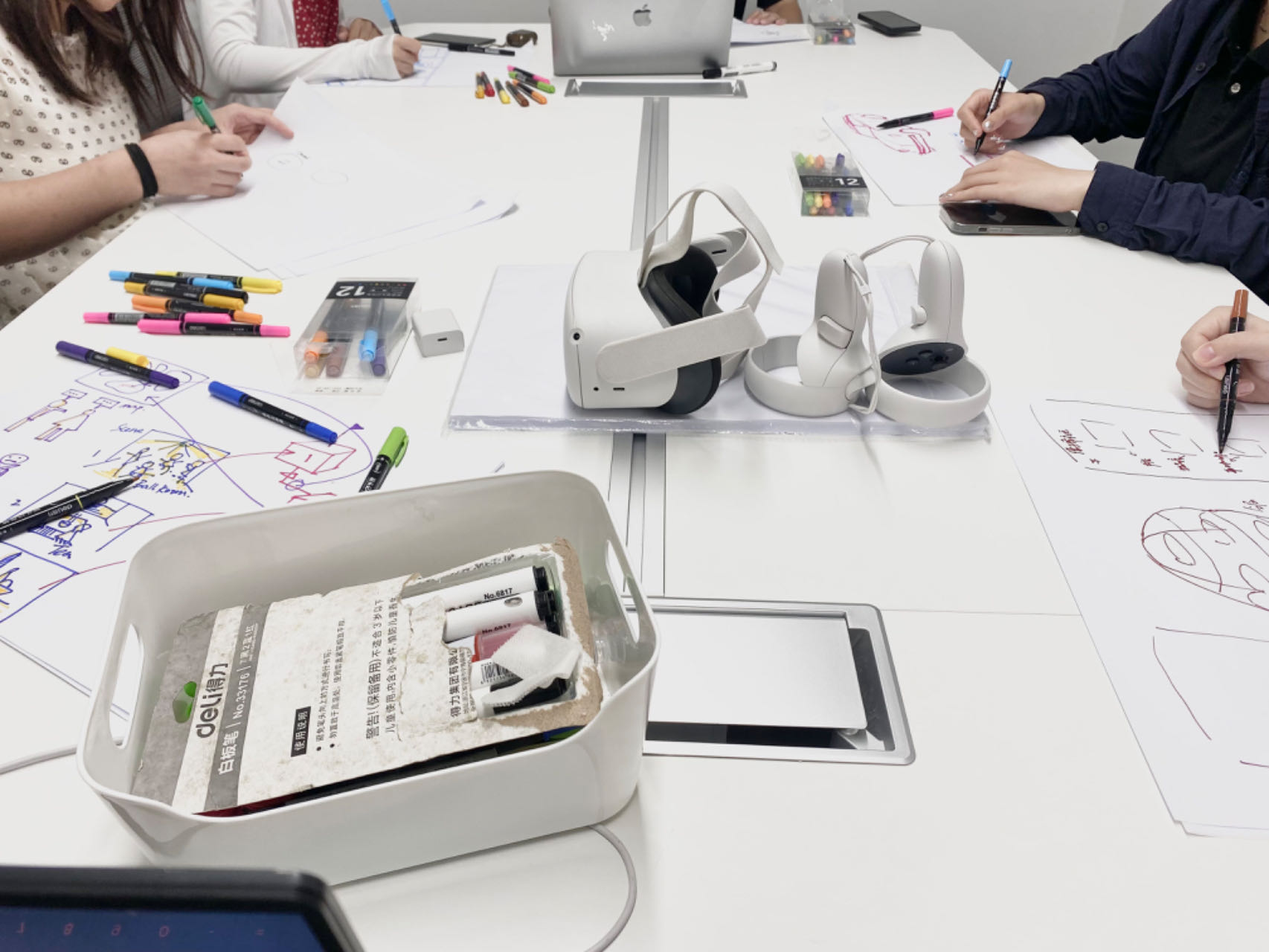} 
        \caption{}
        \label{fig:workshop} 
    \end{subfigure}
    \begin{subfigure}{0.3\textwidth}
        \centering
        \includegraphics[height=2.5cm]{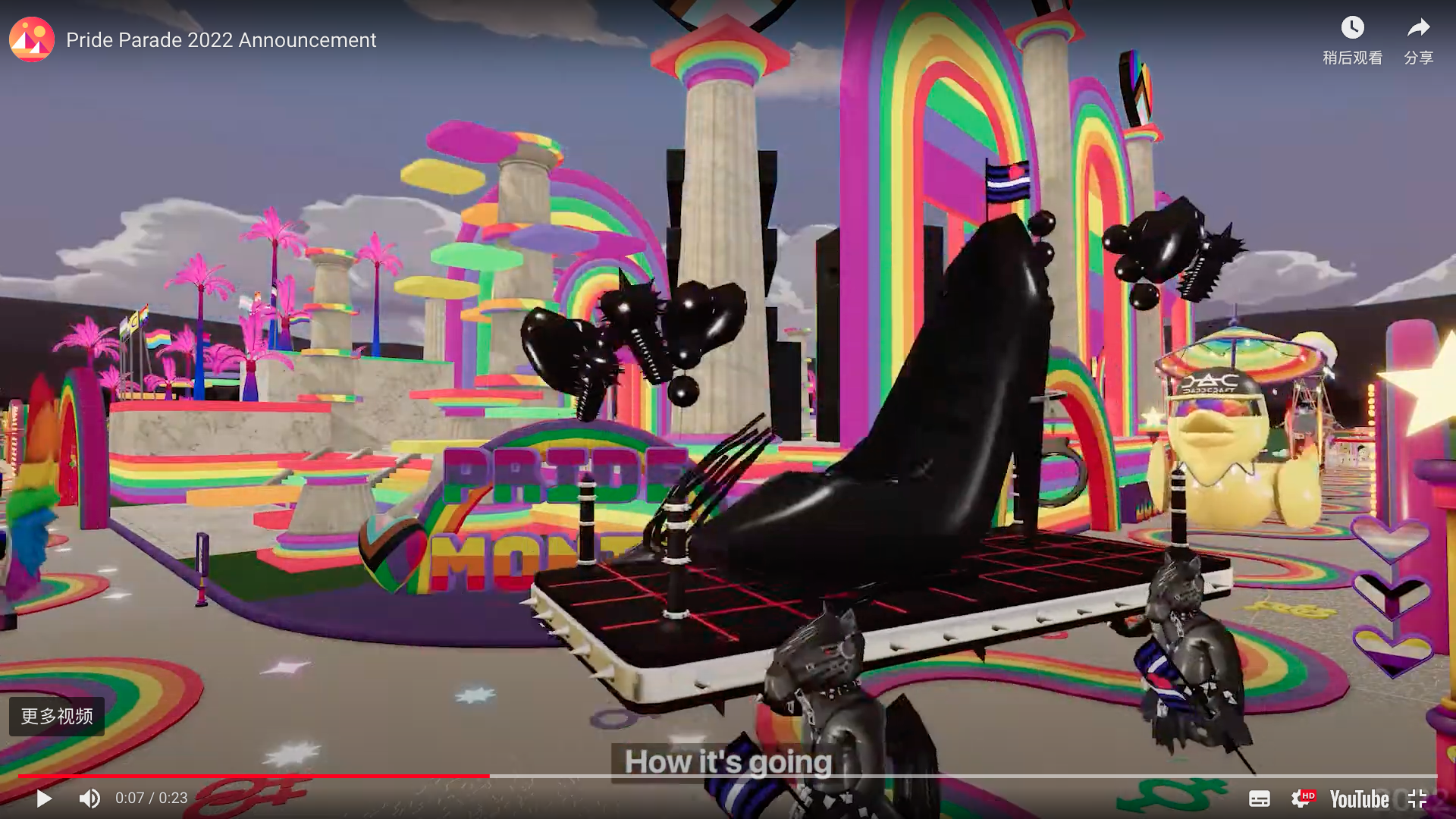}
        \includegraphics[height=2.5cm]{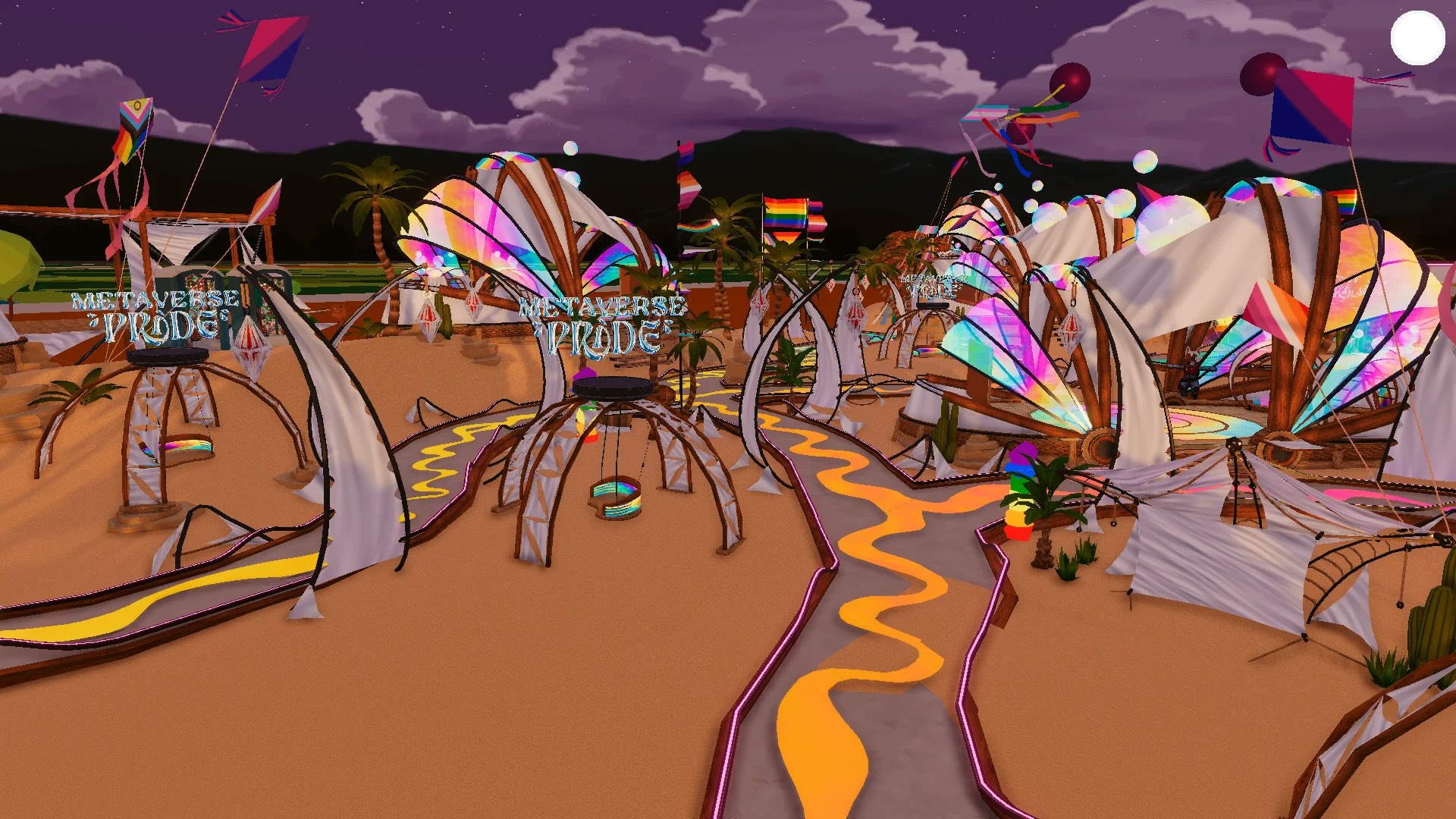}
        \caption{}
        \label{fig:VRscenes}
    \end{subfigure}
    \begin{subfigure}{0.2\textwidth}
        \centering
        \includegraphics[height=5cm]{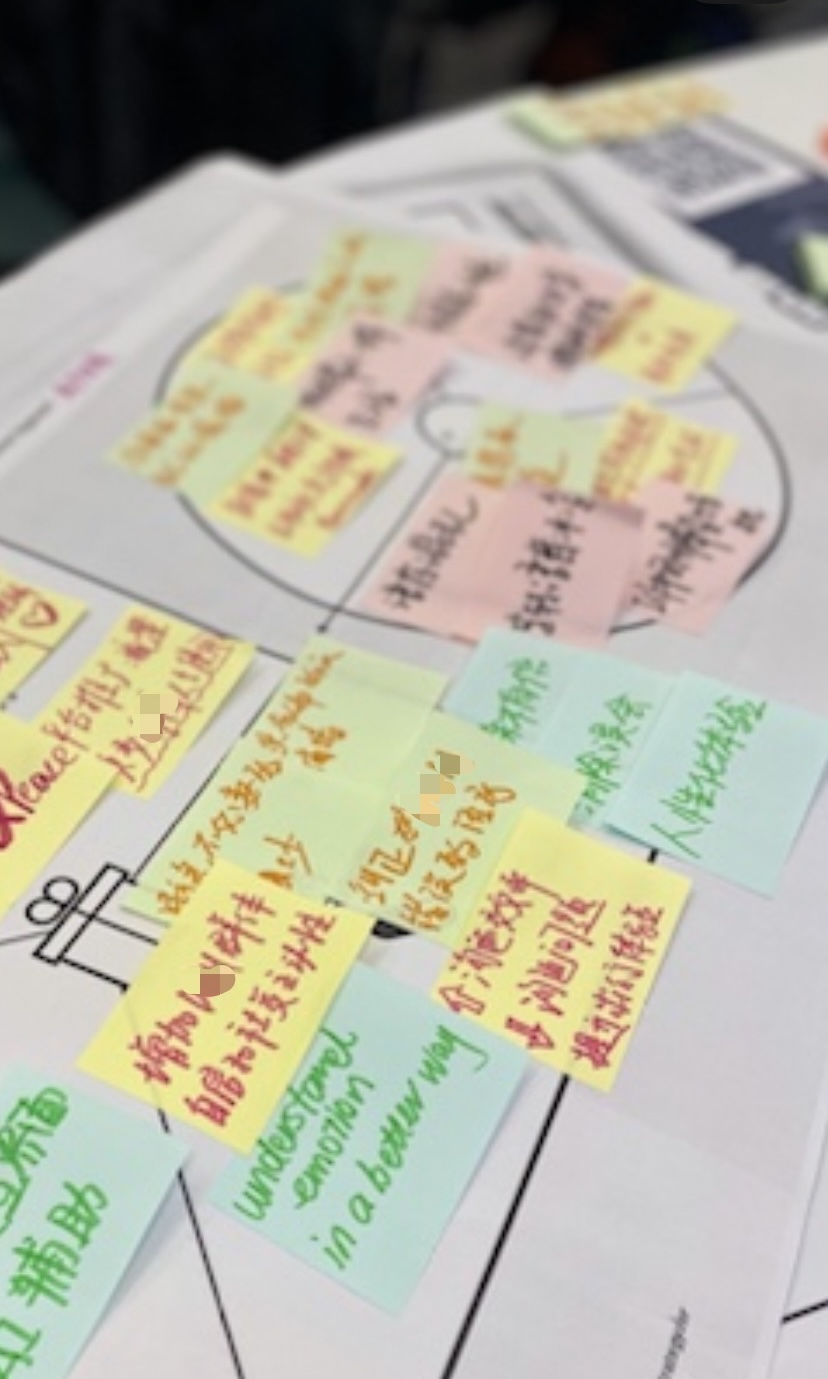}
        \caption{}
        \label{fig:affinitydiagram}
    \end{subfigure}
    \caption{(a) We organized three participatory design workshops involving 18 participants, who contributed rich ideas and virtual sketches based on mutual learning; (b) selected virtual scenarios from LGBTQ+ Pride Month-themed events as part of the ``VR Experience'' phase of preparation; and (c) used an affinity diagram to thematically category communication challenges in the ``Forming Topics'' phase of the study.} 
    \Description{(a) We organized three participatory design workshops involving 18 participants, who contributed rich ideas and virtual sketches based on mutual learning; (b) selected virtual scenarios from LGBTQ+ Pride Month-themed events as part of the ``VR Experience'' phase of preparation; and (c) used an affinity diagram to thematically category communication challenges in the ``Forming Topics'' phase of the study.}
\end{figure*}

\begin{figure*}[ht]
\centering
    \begin{subfigure}[ht]{0.99\textwidth}
        \centering
        \includegraphics[width=\textwidth]{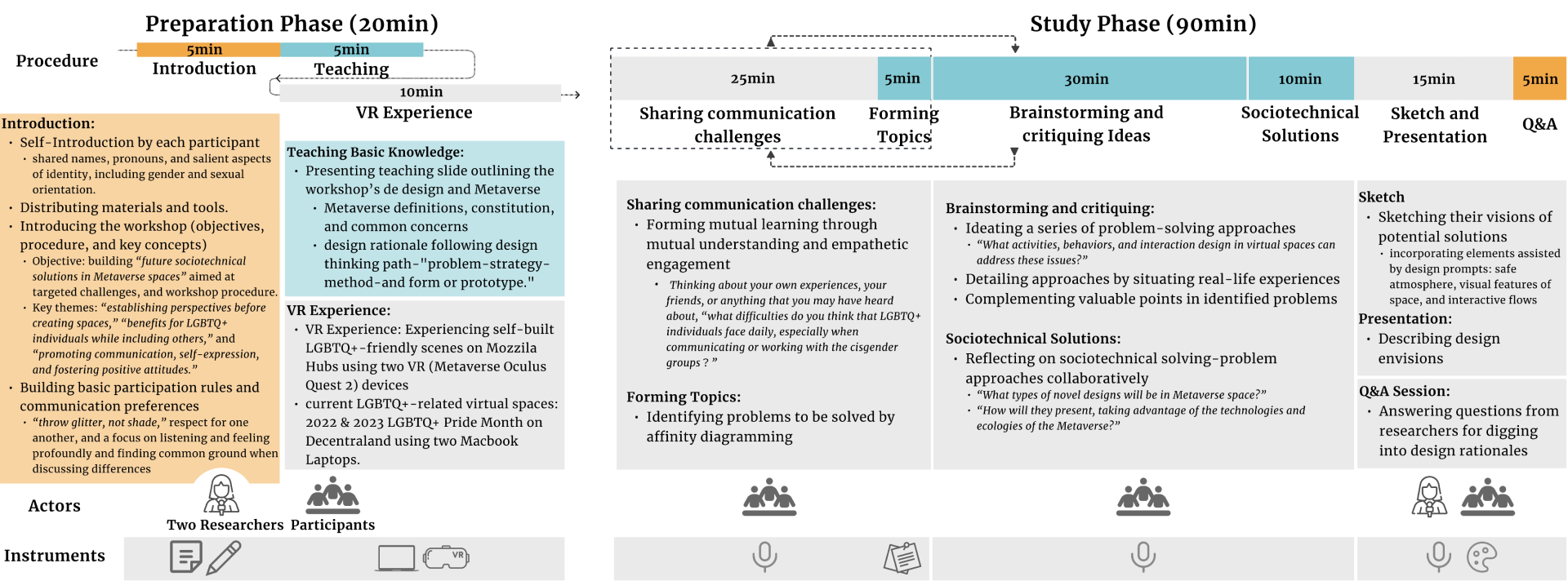}
        \label{fig:workflow}
    \end{subfigure}
    \caption{The workshop flow, including preparation and study phases, and showing each step's details.}
    \Description{The workshop flow, including preparation and study phases, and showing each step's details.}
    \label{fig:workflow}
\end{figure*}

\subsubsection{Protocol Rehearsal}
In the planning phase, we drafted a protocol for future participatory design workshops. To test its feasibility, we organized a rehearsal workshop with our research team, comprising two straight, cisgender women, one straight, cisgender man, and one gay, cisgender man. The rehearsal aimed to evaluate whether the workshop design could foster mutual understanding and address potential misunderstandings between LGBTQ+ participants and cisgender people. 

\subsubsection{Future Workshop}

The workshop flow is illustrated in Figure \ref{fig:workflow}. Before formally starting, each participant introduced themselves by sharing their preferred names, pronouns, and salient aspects of their identity, including gender identity and sexual orientation. Participants were provided with an introductory paper outlining the study’s goals and procedures, along with A3 sheets, pens, markers, and stickers in various colors (Figure \ref{fig:workshop}). 

\paragraph{Introduction}
Next, we presented slides outlining the workshop’s objectives, focusing on exploring ``future socio-technical solutions in Metaverse space,'' to address communication challenges between LGBTQ+ and cisgender people. Key themes included: ``establishing perspectives before creating spaces,'' ``providing benefits for LGBTQ+ people while including others,'' and ``promoting communication, self-expression, and fostering positive attitudes.'' This introduction was followed by basic participation rules and preferences (e.g., ``throw glitter, not shade,'' respect for one another, and focus on listening, feeling profound, and finding common ground when discussing differences). 

We then delivered a brief, 5-minute teaching slide to verbally introduce the Metaverse, covering its definitions, components, common concerns, and design rationale using a design thinking path—
``problem-strategy-method-form or prototype.'' 
\paragraph{VR Experience}
Following this introduction, participants were invited to use two Meta Oculus Quest 2 devices for VR experiences on Mozilla Hubs and two Macbooks to explore Decentraland's 2022 and 2023 LGBTQ+ Pride Month-themed events~\footnote{Metaverse Pride 2022 on Decentraland, guidance and introduction~\url{https://decentraland.org/events/event/?id=5a29fd76-10d7-4879-bb6d-bf68eddb3341}, entry: ~\url{https://decentraland.org/blog/announcements/Metaverse-pride-in-decentraland}.}~\footnote{Metaverse Pride 2023 on Decentraland, ``The Ultimate Guide to Metaverse Pride 2023'', ~\url{https://decentraland.org/blog/announcements/the-ultimate-guide-to-Metaverse-pride-2023}, and entry: ~\url{https://decentraland.org/events/event/?id=1bc791d2-b07f-4857-affd-068bd8b8c992}.} (selected scenes shown in Figure~\ref{fig:VRscenes}) to align with basic knowledge about virtual space and current LGBTQ+ events. 
These activities aimed to familiarize participants with virtual spaces and current LGBTQ+ events. 

\paragraph{Design Stage (1): Identifying Challenges}
Following this, participants were encouraged to reflect on communication challenges faced by LGBTQ+ people when interacting with cisgender people, with questions such as: ``What difficulties do LGBTQ+ people face in daily communication or work with cisgender people?''
Participants took turns sharing personal stories, recording challenges on sticky notes, and collaboratively organized challenge topics using affinity diagramming~\cite{dam2020affinity} after sharing (as shown in Figure~\ref{fig:affinitydiagram}). 

\paragraph{Design Stage (2): Speculative Problem-Solving}
We then encouraged participants to reflect on the potential of Metaverse to address these communication challenges by asking, ``Do you think virtual reality can help solve these communication challenges?'' Based on their experiences and observations, participants brainstormed ideas about activities, behaviors, and interaction designs in virtual spaces that could address these issues. 
During this stage, participants contributed ideas grounded in real-life experiences and continuously added them to sticky notes. 
Next, participants were asked to contemplate socio-technical solutions, focusing on the question: ``What types of novel designs could be implemented in the Metaverse to address these challenges, leveraging its technologies and ecosystem?'' 

\paragraph{Design Stage (3): Visualizing Futures}
Finally, participants sketched their visions of potential solutions, incorporating elements guided by design prompts (e.g., safe atmospheres, visual features of space, and interactive flows) based on their previous discussions. 
Each participant then presented their sketches, explained their ideas, and responded to questions from the researchers. 

\subsection{Data Collection and Analysis}
All workshop dialogues were recorded and transcribed. Following McDonald et al.~\cite{10.1145/3359174} for CSCW and HCI practices, we conducted an inductive thematic analysis to identify concepts, relationships, and patterns among emerging themes. 
This flexible approach provides rich insights, particularly suited for exploring relatively uncharted topics~\cite{guest2011applied, braun2012thematic}. 
Initially, three authors independently coded audio data and physical materials from one workshop (e.g., sketches and notes) using open coding to generate initial codes from the data~\mbox{\cite{charmaz2014constructing, saldana2021coding}}. Through iterative coding, we grouped similar codes (e.g., ``personal shares'' and ``education modules'') into second-level themes, such as "designs to enhance mutual understanding.'' 
Next, we applied axial coding~\mbox{\cite{charmaz2006constructing}} to cluster the data into broader themes after discussion among the three authors. This process resulted in four main themes: ``activities and communication,'' ``interaction features,'' ``collaborative build,'' and ``environmental reform.''

To ensure validity and completeness, two workshop-participating authors independently reviewed the results, addressing overlaps and gaps to reach a consensus. We then held a co-author session to collectively assess and confirm the themes, ensuring a shared understanding.
This iterative process identified four primary themes: ``activities and communication,'' ``interaction features,'' ``collaborative build,'' and ``environmental reform.'' A fifth theme, ``challenges,'' emerged mainly through collaborative affinity diagramming during the design sessions but was not included among the main four.

\subsection{Positionality Statement}
\hl{
In this qualitative research on communication between LGBTQ+ communities and cisgender people within the Metaverse, it is both important and ethical to acknowledge how the researchers' identities and cultural backgrounds may influence the research process, as well as the analysis and interpretation of the data~\mbox{\cite{10.1145/3025453.3025766,10.1145/3443686}}. 
Such disclosure provides transparency by clarifying the researchers' positions in the world, their intellectual and political beliefs, and their goals in conducting the study~\mbox{\cite{renn2010lgbt, gibson2013finding}}. In this study, our team of seven researchers included two LGBTQ+ members who motivated our study and actively participated in all workshops. Each workshop included at least three researchers, enabling diverse perspectives to inform observations and interpretations. All authors have extensive experience as HCI and virtual reality researchers. Our collective identities and experiences helped heighten our awareness of the unique challenges faced by non-cisgender individuals as a marginalized group during our study. 
}

\section{Findings}
    \hl{
In this section, we first outline the two-sided communication challenges identified by our participants. Then, we present the solutions they envisioned through brainstorming and sketching to address some of these issues. The proposed solutions encompass four main design strategies: interactive communication for inclusiveness (Section~\mbox{\ref{sec:4.2communication}}), dynamic social features that balance privacy and openness (Section~\mbox{\ref{sec:socialfeatures}}), scenes and shared experiences to foster intimacy and solidarity (Section~\mbox{\ref{sec:4.4co-futurescenes}}), and unconventional, democratic spaces (Section~\mbox{\ref{sec:4.5reformingnorms}})-corresponding to four layers (as Illustrated in Figure~\mbox{\ref{fig:teaser}}). 
}

\hl{
During the first design session, participants identified several key communication barriers between LGBTQ+ and cisgender individuals through sharing personal experiences. Using collaborative affinity diagramming—a technique for clustering related findings~\mbox{\cite{dam2020affinity}}—we and participants organized these challenges on eight broad categories, as summarized in Table~\mbox{\ref{tab:communication_challenges}}. 
}
\hlorange{
While previous research has extensively documented the challenges LGBTQ+ individuals face (e.g., ~\mbox{\cite{10.1145/3613904.3642494,spiel2019queer}}), mapping these communication barriers into the future Metaverse designs is novel.

Guided by these overarching themes of communication barriers, participants sketched a vision of a future Metaverse that addresses critical issues, including the lack of respect for one’s identity, insufficient safety measures, restricted societal access, and exclusionary social norms. 
These insights serve as the foundation for developing design solutions. 
Although the envisioned solutions did not necessarily target the most pressing challenges facing the community, they were more adaptable to Metaverse-related design considerations than issues such as violence or financial barriers. 
Below, we present the ideas and designs participants developed to address some of the challenges encountered by LGBTQ+ and cis-gender communities. 

}
\begin{table*}[t]
\small
\centering
\arrayrulecolor{white} 
\arrayrulewidth=0.8mm 
\caption{\hl{Summary of communication challenges between LGBTQ+ and cisgender people.}}
\Description{Summary of communication challenges between LGBTQ+ and cisgender people.}
\label{tab:communication_challenges}
\begin{tabular}{|>{\centering\arraybackslash}m{0.8cm}|>{\raggedright\arraybackslash}p{4cm}|>{\raggedright\arraybackslash}p{9cm}|}
\hline
\multirow{5}{*}{\parbox[c]{1cm}{\centering\rotatebox{90}{\parbox[c]{3cm}{\centering\textbf{Challenges for Cisgender Group}}}}}&\cellcolor{gray!30}\textbf{Challenge Type} & \cellcolor{gray!30} \textbf{Description}\\
\cline{2-3}
& \cellcolor{lightgray!10} \textbf{1. Lack of Awareness} & \cellcolor{lightgray!10} Limited exposure to LGBTQ+ individuals in their social circles. \\
\cline{2-3}
& \cellcolor{gray!15} \textbf{2. Fear of Misunderstanding} & \cellcolor{gray!15} Hesitation to engage with LGBTQ+ individuals due to societal misconceptions and the influence of public opinion. \\
\cline{2-3}
& \cellcolor{lightgray!10} \textbf{3. Concerns About Acceptance} & \cellcolor{lightgray!10} Anxiety about facing unwelcoming attitudes when interacting with LGBTQ+ individuals. \\
\cline{2-3}
& \cellcolor{gray!15} \textbf{4. Misconceptions and Stigma} & \cellcolor{gray!15} Fear of unintentionally offending LGBTQ+ individuals or perpetuating stereotypes. \\
\midrule
\multirow{6}{*}{\parbox[c]{1cm}{\centering\rotatebox{90}{\parbox[c]{3cm}{\centering\textbf{Challenges for LGBTQ+ Group}}}}}
&  \cellcolor{orange!20} \textbf{1. Misunderstandings and Stereotypes} & \cellcolor{orange!20} Dealing with and correcting misconceptions, including harmful stereotypes such as the ``trap'' narrative.\\
\cline{2-3}
&  \cellcolor{orange!10} \textbf{2. Fear of Judgment} & \cellcolor{orange!10} Anxiety about being judged or negatively perceived by others. \\
\cline{2-3}
&  \cellcolor{orange!20} \textbf{3. Safety Concerns} & \cellcolor{orange!20} Concerns about personal safety based on previous experiences in digital spaces, such as social media or virtual reality. \\
\cline{2-3}
&  \cellcolor{orange!10} \textbf{4. Information Interchange} & \cellcolor{orange!10} Difficulty in explaining LGBTQ+ identities and experiences to others. \\
\cline{2-3}
&  \cellcolor{orange!20} \textbf{5. Intersectional Challenges} & \cellcolor{orange!20} Struggles in explaining sexual identity and orientation, particularly transgender identities, even within LGBTQ+ communities. \\
\cline{2-3}
&  \cellcolor{orange!10} \textbf{6. Lack of Trust} & \cellcolor{orange!10} Difficulty building trust, often due to past harmful interactions. \\
\hline
\end{tabular}
\end{table*}

\subsection{Interactive Communication with High Inclusiveness}~\label{sec:4.2communication}
 All participants emphasized the lack of contact and the prevalence of misconceptions (as shown in Table~\mbox{\ref{tab:communication_challenges}}) posed significant barriers to effective communication. They proposed various design ideas for collaborative activities, avatar features, and multi-spaces for more equitable communication between LGBTQ+ and cisgender people. 

    
\subsubsection{Designing for Meaningful Interactive Communication.}~\label{sec:4.2communication}
These proposals underscore the importance of activities that challenge and transcend heteronormative assumptions, thereby fostering more meaningful communication. All participants highlighted the necessity of being recognized for their complete identity rather than being solely defined by their gender. 

   \textbf{Activities with Embodied Interaction.} 
    Participants proposed gathering activities
     using rich embodied interactions to promote communication. 
    These interactions involve nuanced expressions of body language, gestures, and facial cues. For example, participants mentioned activities such as \textit{``taking group photos''} (P2, Cis Male, Heterosexual), \textit{``drinking tea while casually observing''} (P4, Cis Male, Gay), and \textit{``baking around a campfire''} (P5, Cis Male, Heterosexual). Through full-body embodied perception, these activities facilitate effective and expressive communication. As P10 (Cis Male, Heterosexual) reflected, \textit{``through these activities, they are just friends to me—living individuals, not marginalized ones.''} 
    Additionally, participants envisioned \textit{``virtual special effects''} (digital enhancements or visual modifications) as personalized embodiment features that transcend binary gender representations. For example, P13 (Cis Female, Heterosexual) proposed visual indicators of hobbies, such as \textit{``gleamy icons''} (dynamic, glowing symbols that represent users' interests) appearing around users’ feet while walking, designed to \textit{``help others identify shared interests and connect organically''} (shown in Figure~\ref{fig:PP13}). Such features could foster connections based on shared interests, encourage natural conversations, and help form inclusive communities.

    \textbf{Diverse and Composite Spaces.} 
    Participants envisioned the future Metaverse as a composite environment that includes diverse spaces to promote inclusiveness, where \textit{``daily life, events, and parties can all find a place here''} (P15, Cis Female, Lesbian). 
    In these composite spaces, participants envisioned a coexistence of seemingly opposing and stylistically diverse activities, using a space-switching mode, switching between \textit{``academic education''} and \textit{``virtual parade''} (P10, Cis Female, Heterosexual), as well as between \textit{``catwalk show''} and \textit{``solemn performance''} (P9, Non-Binary, Pansexual). 
    Such coexistence would create dynamic interactions and layered experiences to enhance inclusiveness and creativity. Together, these features envision the Metaverse as a versatile environment where diverse needs and perspectives are integrated. 


\subsubsection{Designing for Enhanced Mutual Understanding.}~\label{sec:educationmodules}
In addition to seeking increased opportunities for interaction, many participants expressed a need to address misunderstandings and misconceptions about LGBTQ+ groups. They believe that immersive embodied experiences offer a more effective means of achieving this compared to conventional digital communities. 
 
\begin{figure}[t]
\centering
    \begin{subfigure}[t]{0.49\textwidth}
    \centering
    \includegraphics[height=5.5cm]{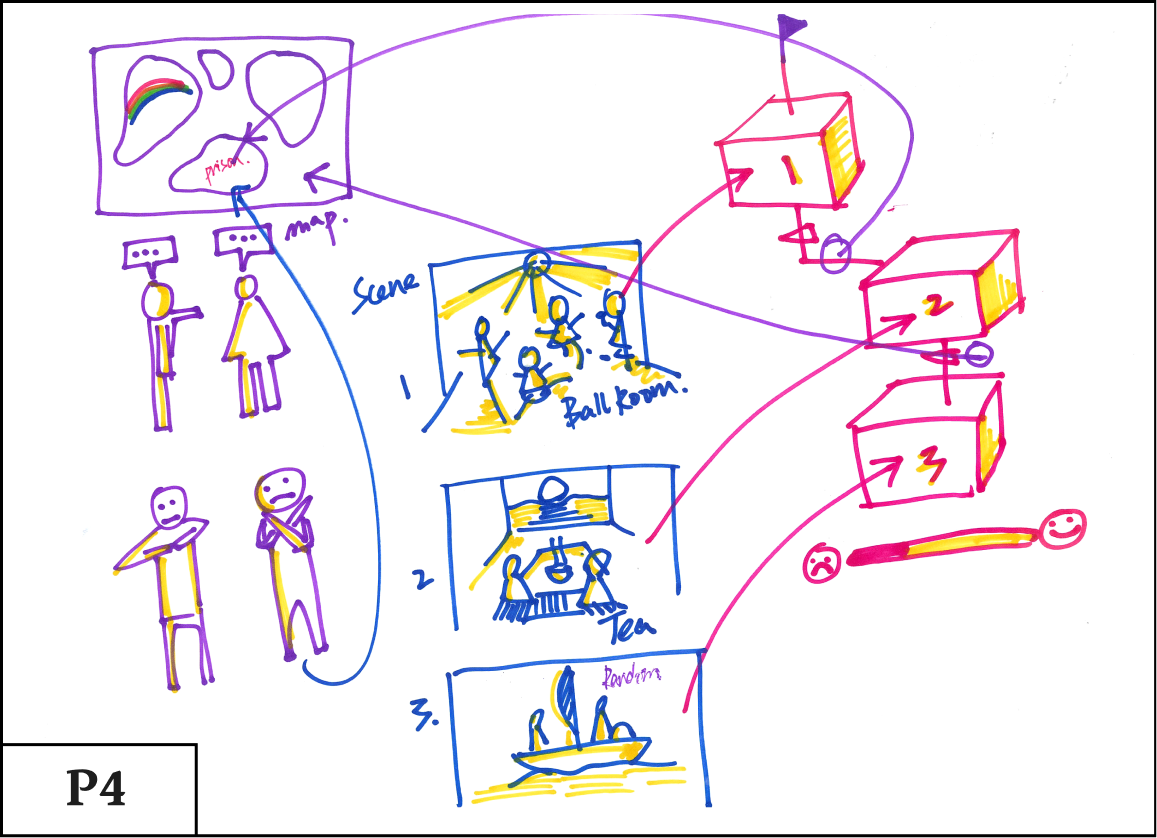}
    \end{subfigure}
\caption{An island adventure combined with a narrative experience as a game (Workshop 1, P4).}
\Description{An island adventure combined with a narrative experience as a game (Workshop 1, P4).}
    \label{fig:PP4}
\end{figure}

 \textbf{Inclusive Education and Awareness Building.} 
Participants envisioned immersive educational modules that contains diversified elements. First, an \textit{``onboard registration process would introduce users to the basics of gender and identity,''} providing essential background knowledge. This foundational understanding would then be tested and reinforced through immersive, \textit{``quick quizzes embedded within simulated events.''} In these scenarios, users would be required to assess whether certain words and actions align with principles of inclusiveness. As one participant described, \textit{``the quizzes would ‘test users’ responses in situational scenarios to explain platform rules,''} offering a dynamic and engaging method for both educating and assessing cultural sensitivity in real-world contexts'' (P8, Trans Female, Lesbian). 
Participants also envisioned involving a \textit{``point system''} in these simulated educational scenarios to reward user long-term learning, such as \textit{``a series of lectures as open sources''} (P3, Cis Female, Lesbian). This system could also hold users accountable for being unfriendly and violent towards others using ubiquitous data. \textit{``Users are scored on actions and interactions, with points unlocking advanced features. Otherwise, they'll be sent to a Lonely Island once their score falls below the basic threshold''} (P4, Cis Male, Gay; visualized an event in Figure \ref{fig:PP4}). These suggestions underscore the necessity to cultivate a more responsible and supportive virtual community culture. 

\begin{figure}[h]
\centering
    \begin{subfigure}[h]{0.49\textwidth}
    \centering
    \includegraphics[height=5.5cm]{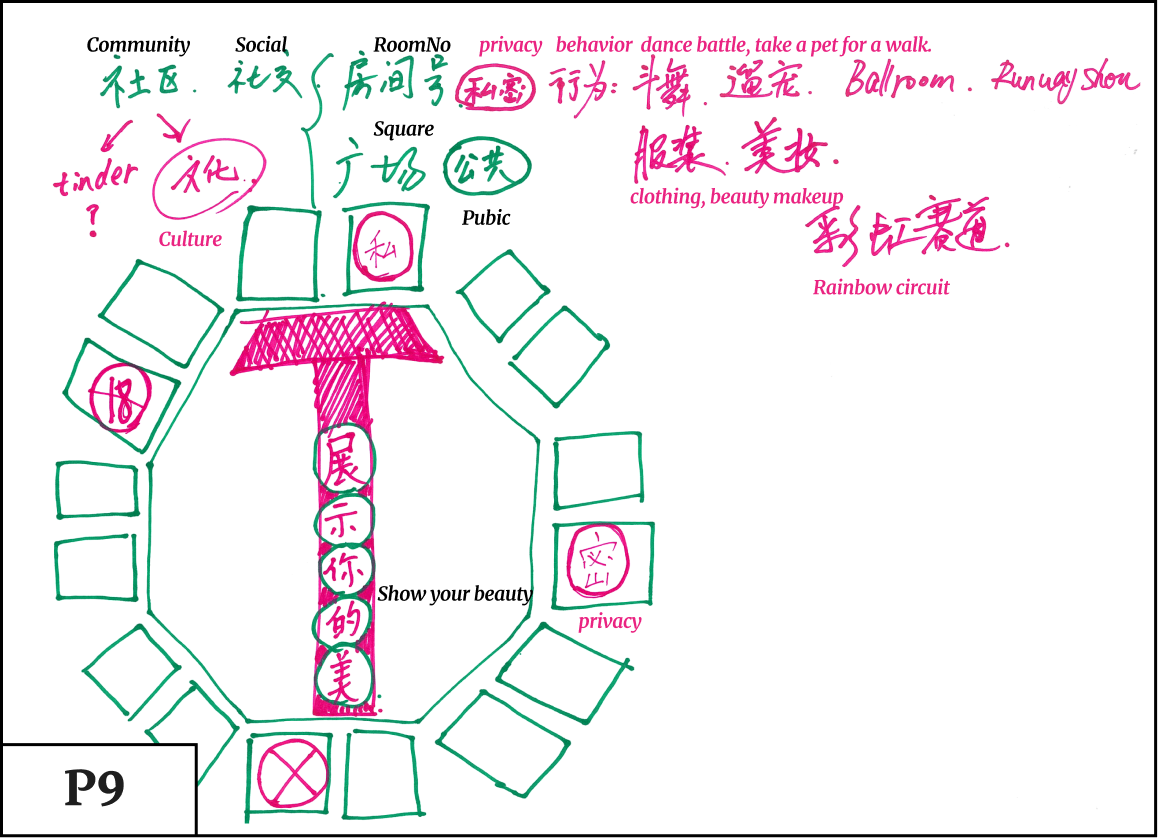}
    \caption{}
    \label{fig:PP9}
    \end{subfigure}
    \begin{subfigure}[h]{0.49\textwidth}
    \centering
    \includegraphics[height=5.5cm]{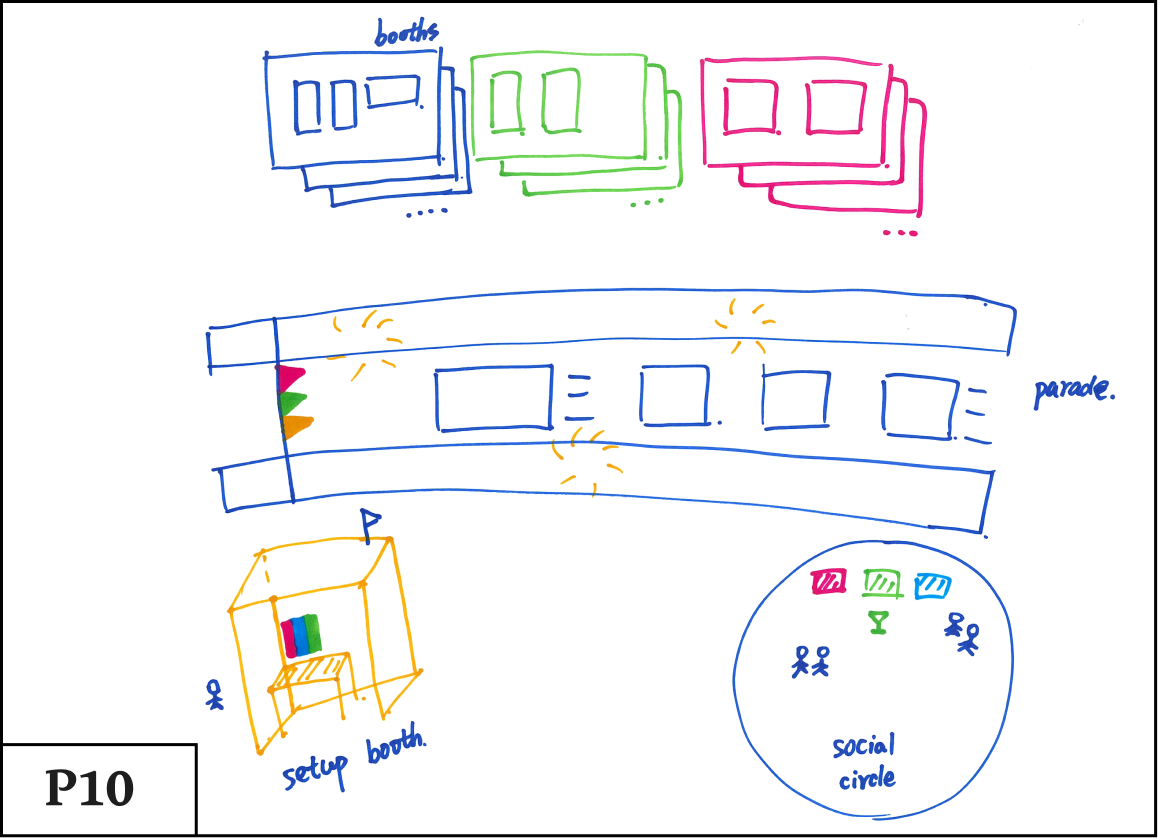}
    \caption{}
    \label{fig:PP10}
    \end{subfigure}
\caption{(a) A runway for showcasing beauty, surrounded by semi-private rooms for discussing more private topics (Workshop 2, P9); (b) Outdoor Metaverse spaces offering options like roadshows and pride parades (Workshop 2, P10).}
\Description{(a) A runway for showcasing beauty, surrounded by semi-private rooms for discussing more private topics (Workshop 2, P9); (b) Outdoor Metaverse spaces offering options like roadshows and pride parades (Workshop 2, P10).}
\end{figure}

\textbf{Cultural Celebration and Representation.} 
As participants' enthusiasm for \textit{``bringing LGBTQ+ subcultures into public view,''} many envisioned LGBTQ+ events, such as virtual Pride parades, catwalks, and street culture representations. P9 (Non-Binary, Pansexual) designed a ballroom-inspired catwalk called \textit{``Showing your Beauty''} where \textit{``users could showcase their appearance and receive comments''} (visualized in Figure~\ref{fig:PP9}). These celebrations and representations could significantly support users living in regions where LGBTQ+ communities have been silenced or marginalized; as P10 (Cis Female, Heterosexual) mentioned, \textit{``The future Metaverse should allow these repressed cultures to thrive''} (visualized in Figure~\ref{fig:PP10}). 
Participants also envisioned various promotional strategies using celebrity influence for LGBTQ+ communities, including keynote speeches, lectures, and film festivals. P18 envisioned \textit{``cartoonish avatars of LGBTQ+ celebrities''} into the Metaverse, such as \textit{``abstract representations of figures like Cook or Turing''} (P18, Cis Male, Unsure; visualized this idea in Figure~\ref{fig:PP18}). This lighthearted and symbolic representation could enhance accessibility and relatability, serving as a welcoming and inclusive symbol within the Metaverse. 
These envisioned designs reflect participants' aspirations for challenging stereotypes and symbolizing a more accepting society by putting LGBTQ+ people in the social spotlight. 
    

\begin{figure}[h]
\centering
    \begin{subfigure}[t]{0.49\textwidth}
    \centering
    \includegraphics[height=5.5cm]{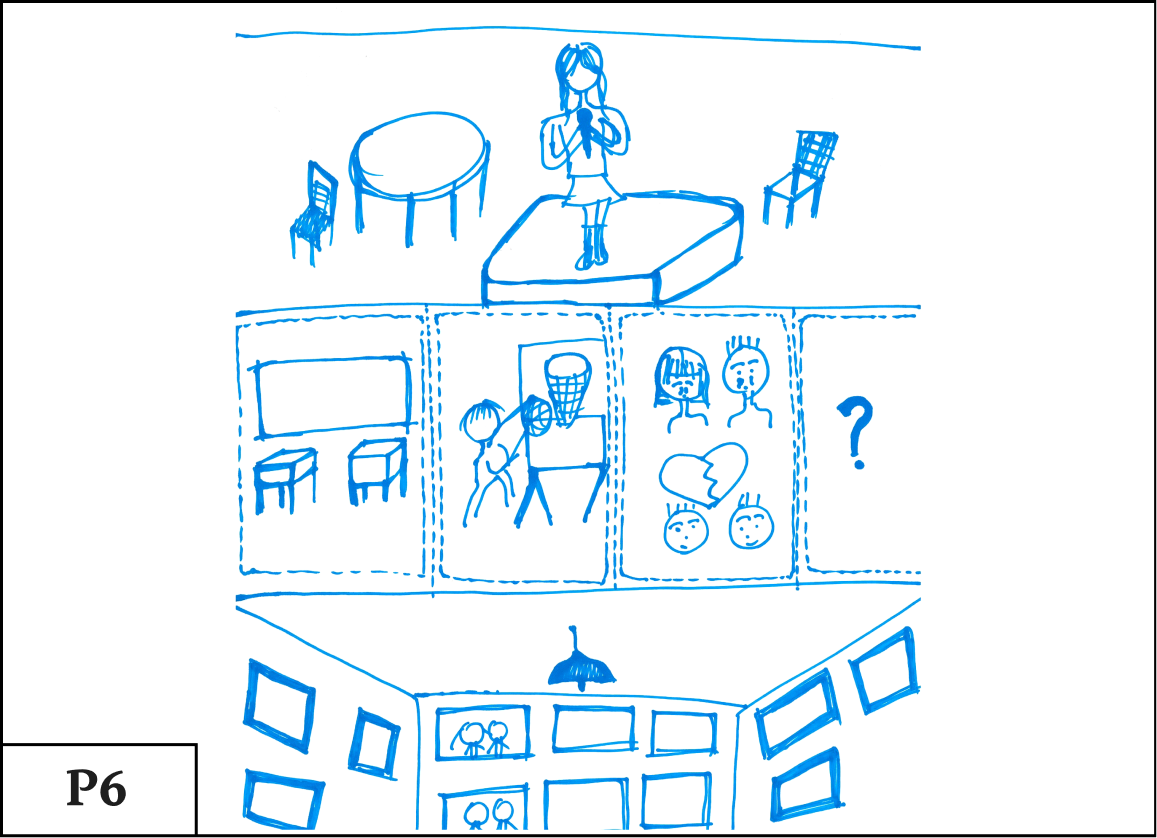}
    \end{subfigure}
\caption{A community building creating a ``community'' through activities and lives, where one room has a wall filled with memories and commemorative photos representing unique stories (Workshop 1, P6).}
\Description{A community building creating a ``community'' through activities and lives, where one room has a wall filled with memories and commemorative photos representing unique stories (Workshop 1, P6).}
    \label{fig:PP6}
\end{figure}

    \textbf{Empathetic Communication and Personal Connection.}
Several participants proposed designing personal storytelling that utilizes both virtual and physical acting. For instance, P1 (Cis Female, Heterosexual) envisioned a virtual shared channel where \textit{``Individuals from both groups could watch or listen to vloggers narrating real-life experiences, allowing people to understand who LGBTQ+ individuals are, moving beyond stereotyped definitions.''} This approach would encourage genuine engagement by informing authentic identity and fostering empathy through content that challenges preconceived notions. While these ideas were beneficial, participants also acknowledged that \textit{``It can lead to either deeper understanding or deeper hurt, so revealing sincere feelings requires courage''} (P6, Cis Female, Heterosexual). 
To mitigate these emotional risks, P6 (Cis Female, Heterosexual) suggested incorporating asynchronous media, such as \textit{``personal photographs,''} as safer alternatives (shown in Figure~\ref{fig:PP6}). 
Similarly, P1 proposed developing a user-generated content (UGC) video platform featuring a \textit{``bullet screen''} function (a feature that allows users to post real-time, floating comments over a video as it plays), enabling users to interact and explore the experiences of LGBTQ+ people while watching VR vlogs instead of in-person sharing. This feature would facilitate asynchronous, supportive exchanges, allowing users to engage without exposing themselves to unnecessary risk.
These thoughts reflect individuals' desire to gain attention through genuine understanding and accurate awareness, which are considered fundamental to effective communication. 

\subsection{Dynamic Social Features that Balance Privacy and Openness}~\label{sec:socialfeatures}
Our LGBTQ+ participants reported heightened concerns about personal safety and social judgment. Consequently, they emphasized the need for design features that balance safety and interaction by supporting flexible social cues that foster meaningful participation.

\subsubsection{Designing for Flexible Social Interaction.}~\label{sec:flexibleinteraction} 
Participants recommended context-sensitive social cues, enabling to customize their own interactions based on personal comfort levels. This design strategy gives LGBTQ+ users greater control over their virtual experiences. 
    
\textbf{Visualized Intimacy Level.}
Participants envisioned using avatars' appearances, e.g., resolution and colors, to reflect the varying levels of intimacy with ongoing social interactions. As P13 (Cis Female, Heterosexual) described, this system could work similarly to the immersive game \textit{Sky: Children of the Light}: \textit{``Initially, others appear gray and lack detail. As we become closer, their appearance becomes colorful and luminous.''}  This system could provide \textit{``a degree of concealment''} during initial interactions, enabling users to engage while preserving privacy (P14, Cis Female, Bisexual). It could also be regarded as \textit{``a consent mechanism''} for interactions between both choices, as P18 (Cis Male, Unsure) explained, \textit{``If a party in a relationship tend not to close, the intimacy between them will be not high. The other party only sees the avatar with the missing details.''} emphasizing using social cues to promote more inclusive and context-sensitive social exchanges.
These design ideas reflect the importance of adaptive interpersonal boundaries and consensual visual cues in fostering safer, more nuanced, and equitable interactions in the future Metaverse. 

\begin{figure}[h]
\centering
    \begin{subfigure}[t]{0.49\textwidth}
    \centering
    \includegraphics[height=5.5cm]{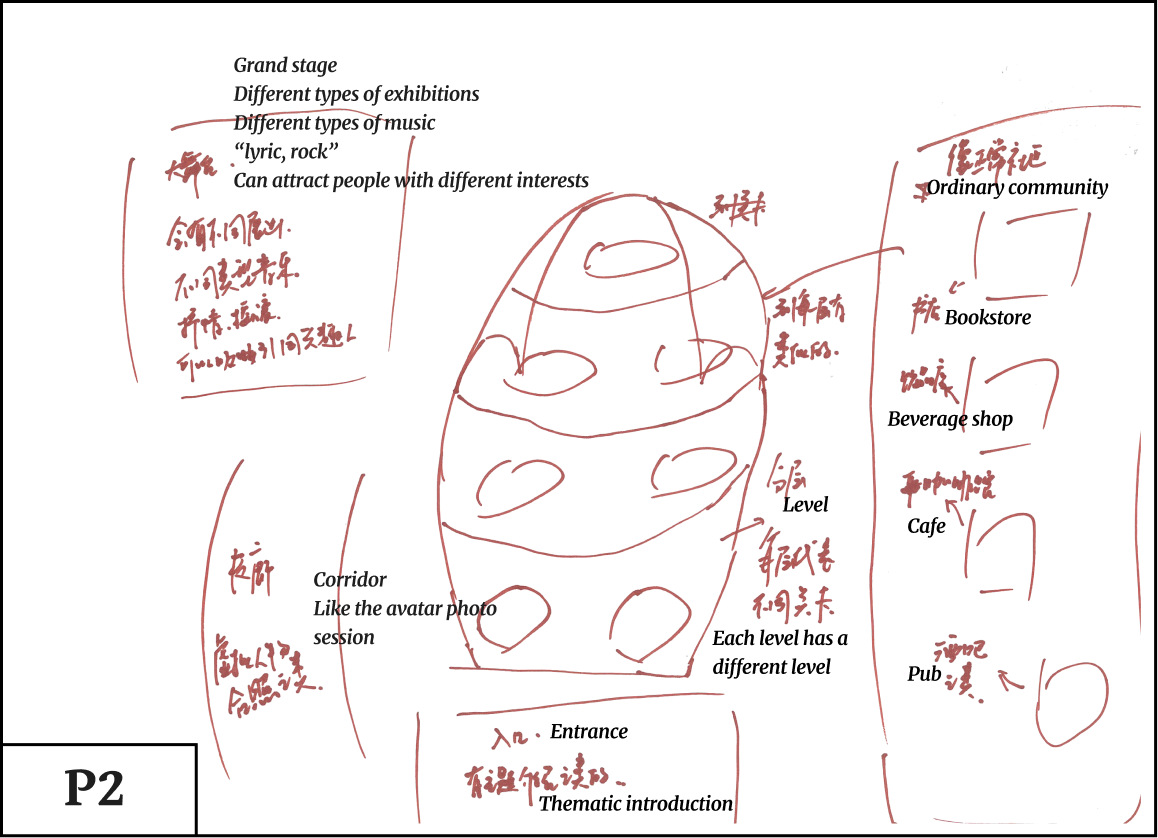}
    \Description{A vertical architectural space, with many levels referring to different areas (Workshop 1, P2).}
    \end{subfigure}
\caption{A vertical architectural space, with many levels referring to different areas (Workshop 1, P2).}
\Description{A vertical architectural space, with many levels referring to different areas (Workshop 1, P2).}
    \label{fig:PP2}
\end{figure}

\textbf{Adaptable Spatial Boundary.}  
Participants suggested another flexible model of social interaction using \textit{``a branching narrative with levels''} to define adaptive spatial boundaries. This system would dynamically adjust based on user behavior and interaction, as envisioned by P8 (Non-Binary, Pansexual): \textit{``High accumulative points in trust level allow advancement to larger event areas, such as bigger and bigger dance floors.''} 
By progressing through levels, users can reduce social distance and engage more closely, \textit{``The larger dance floor that high trust allows access to screen out people who are more willing to interact, and will allow more behaviors to occur, such as hugging, holding hands''} (P8, Non-Binary, Pansexual). This feature could also enhance security by serving as a spatial boundary that guarantees fully informed consent. 
In this regard, several participants also envisioned diverse spatial forms, such as vertical \textit{``hive-like''} and horizontal structures \textit{``stacked boxes.''} For example, P2 (Cis Male, Heterosexual) envisioned \textit{``a large, mountain-like structure with a spiral path''} as level settings where \textit{``users could encounter different activities on LGBTQ+-related topics''} (see Figure~\ref{fig:PP2}). This spatial design strategy provides a multi-sensory experience that engages users within a narrative-based level experience. 


\subsubsection{Designing for Secure Privacy and Safety.}~\label{sec:safeboudary}
Although participants suggested adaptive mechanisms for managing social interaction, some were also concerned about privacy and safety, as stated: \textit{``Somebody may lack self-confidence and trust toward cisgender people due to previous bad experiences''} (P10, Cis Female, Heterosexual). 
Therefore, participants envisioned more safety designs in a fixed way. 

\begin{figure}[h]
\centering
    \begin{subfigure}[t]{0.49\textwidth}
    \centering
    \includegraphics[height=5.5cm]{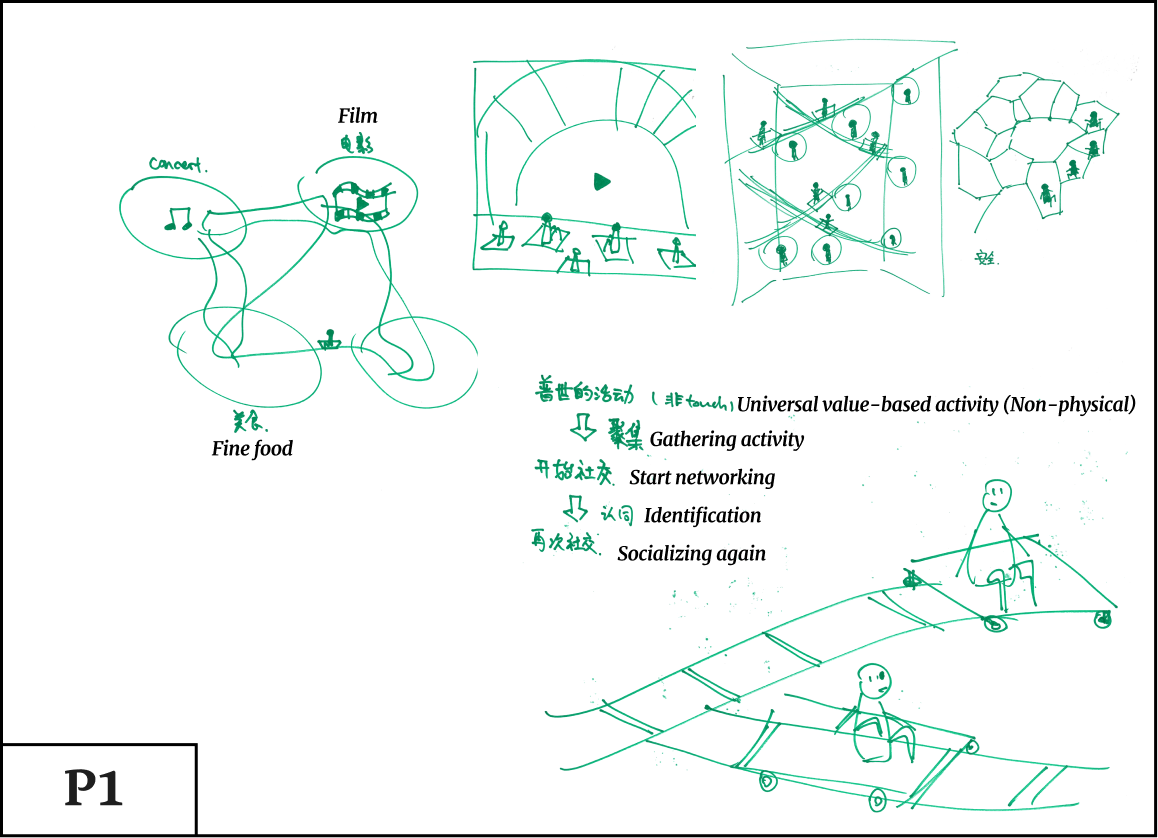}
    \Description{Space boundary designs based on interests and a social interaction model based on track gliding (Workshop 1, P1).}
    \end{subfigure}
\caption{Space boundary designs based on interests and a social interaction model based on track gliding (Workshop 1, P1).}
\Description{Space boundary designs based on interests and a social interaction model based on track gliding (Workshop 1, P1).}
    \label{fig:PP1}
\end{figure}

\textbf{Manageable Personal Boundary.}
Participants envisioned innovative ways to manage personal boundaries through spatial motion and interaction rules. For instance, P1 (Cis Female, Heterosexual) proposed a scene with scooters, where \textit{``users navigate along tracks, encountering others who can choose to either carpool or continue independently.''} This pairing mechanism would \textit{``ensure personal boundaries by balancing comfort levels and interaction opportunities''} (P1, Cis Female, Heterosexual; visualized this idea in Figure~\ref{fig:PP1}). Such designs could minimize harmful interactions within personal networks, enhancing privacy and security in immersive embodied experiences. 
In addition, the concept of a \textit{``social bubble''} emerged as a critical feature for boundary protection. This bubble would \textit{``prevent physical contact by maintaining a minimum distance and include a visual communication interface to display individual social distances within the Metaverse.''} Additionally, it could \textit{``trigger alarms or emergency mechanisms, repelling an aggressor if someone violates your space or body within the bubble''} (P1, Cis Female, Heterosexual). These measures reflect the need for real-time, adaptable security systems responsive to user behavior. 

\textbf{Fixed Spatial Boundary.} 
In addition to dynamic boundary management, participants proposed fixed spatial boundaries, for instance, using \textit{``portals to transition between scenes''} enabling invitation-only access (P9, Non-Binary, Pansexual). This feature could offer varying levels of privacy depending on the activity, as P9 explained: \textit{``There are rooms that are 18+ restricted, while others are semi-private, with different permissions for access''} (visualized this idea in Figure~\ref{fig:PP9}). Such an approach allows users to select spaces that align with their privacy preferences, enhancing safety and comfort. 
Participants also envisioned exclusive personal spaces within the future Metaverse. P1 (Cis Female, Heterosexual) imagined \textit{``several exclusive paths to access my closest friend's house,''} where fixed zones could serve as personal sanctuaries, limiting interactions to invited guests (see Figure~\ref{fig:PP1}). These spaces would shield individuals from unwanted encounters, providing a secure haven where individuals can feel at ease.

\subsection{Scenes and Shared Experiences with Intimacy and Solidarity}~\label{sec:4.4co-futurescenes}
Participants proposed strategies that particularly address facilitating relaxing and intimate interactions, as well as fostering community solidarity through co-creation. These strategies aim to build deep connections between individual presence and the Metaverse

\subsubsection{Designing for Personal Intimate Connection.}
\label{sec:nuancedsocialcues} 
To address the challenges of the fear of misunderstanding and nonacceptance, participants proposed spaces and scenes that foster relaxation and intimacy as light-touch strategies. 

\begin{figure}[h]
\centering
    \begin{subfigure}[t]{0.49\textwidth}
    \centering
    \includegraphics[height=5.5cm]{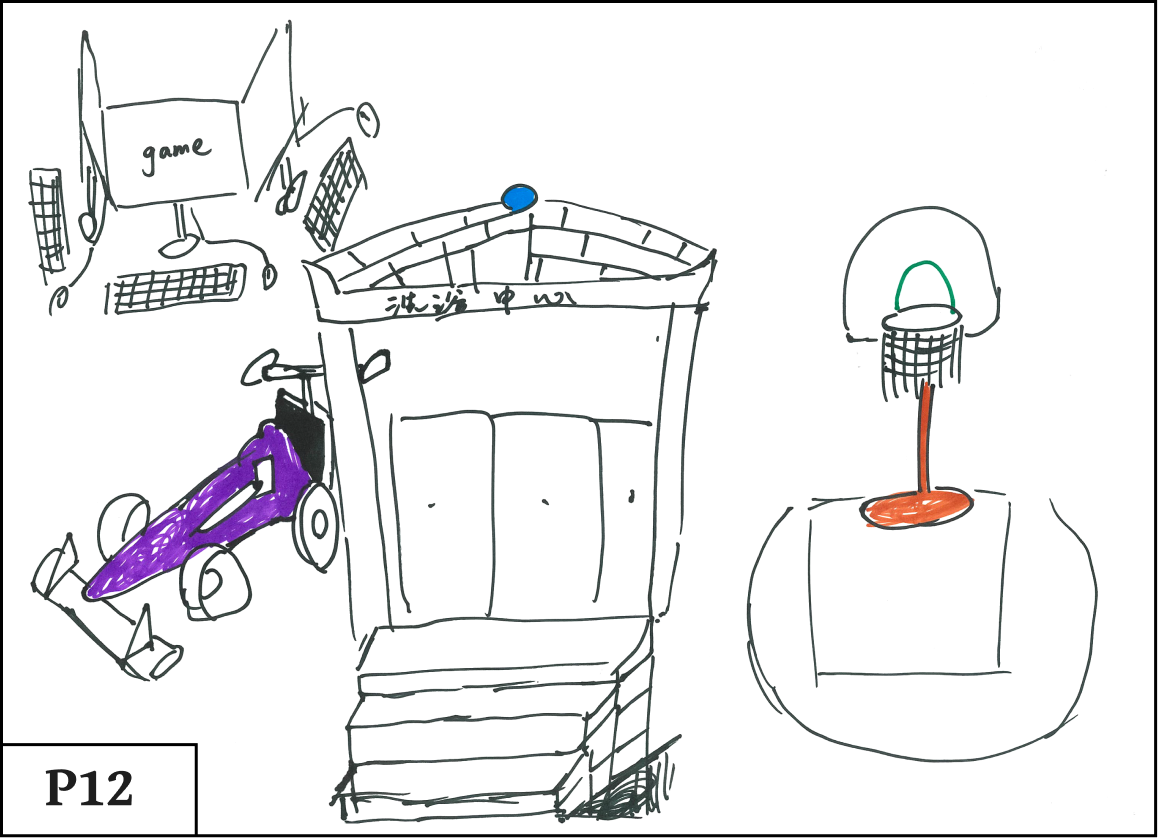}
    \end{subfigure}
\caption{A shared space for sports and lifestyle activities (Workshop 2, P12).}
\Description{A shared space for sports and lifestyle activities (Workshop 2, P12).}
    \label{fig:PP12}
\end{figure}

\textbf{Inhabit and Living Scene.} 
Participants emphasized non-goal-oriented behaviors, envisioning experiences of inhabiting and living as digital natives, extending beyond conventional virtual interactions. Proposed designs included a \textit{``bath center''} (P12; visualized this idea in Figure~\ref{fig:PP12}), a \textit{``leisure island''} (P4) for relaxation, and mixed reality interactions like \textit{``raising a glass in the physical world while cheering virtually''} (P5, Cis Male, Heterosexual). These designs aim to create unstructured environments that foster organic interactions and promote mental well-being. 
Natural settings were particularly favored, such as \textit{``sea island,''}\textit{``seaside,''} \textit{``tropical rainforest.''} P5 (Cis Male, Heterosexual) described a campfire by the sea as a setting for \textit{``storytelling and emotional bonding.''} While P15 (Cis Female, Lesbian) highlighted the symbolic importance of the sea, referring to it as a \textit{``spiritual Jerusalem''} in the Metaverse, ideal for fostering intimacy. Other participants preferred alternative, escapist worlds, such as a \textit{``less rational''} relaxation space with a \textit{``meditative atmosphere''} (P10, Cis Female, Heterosexual), or a futuristic, colorful world rich in advanced technology, offering spiritual respite (P14, Cis Female, Bisexual). These imaginative designs underscore the importance of diverse, personalized spaces that resonate with participants' emotional and spiritual needs. 

\begin{figure}[h]
\centering
    \begin{subfigure}[t]{0.49\textwidth}
    \centering
    \includegraphics[height=5.5cm]{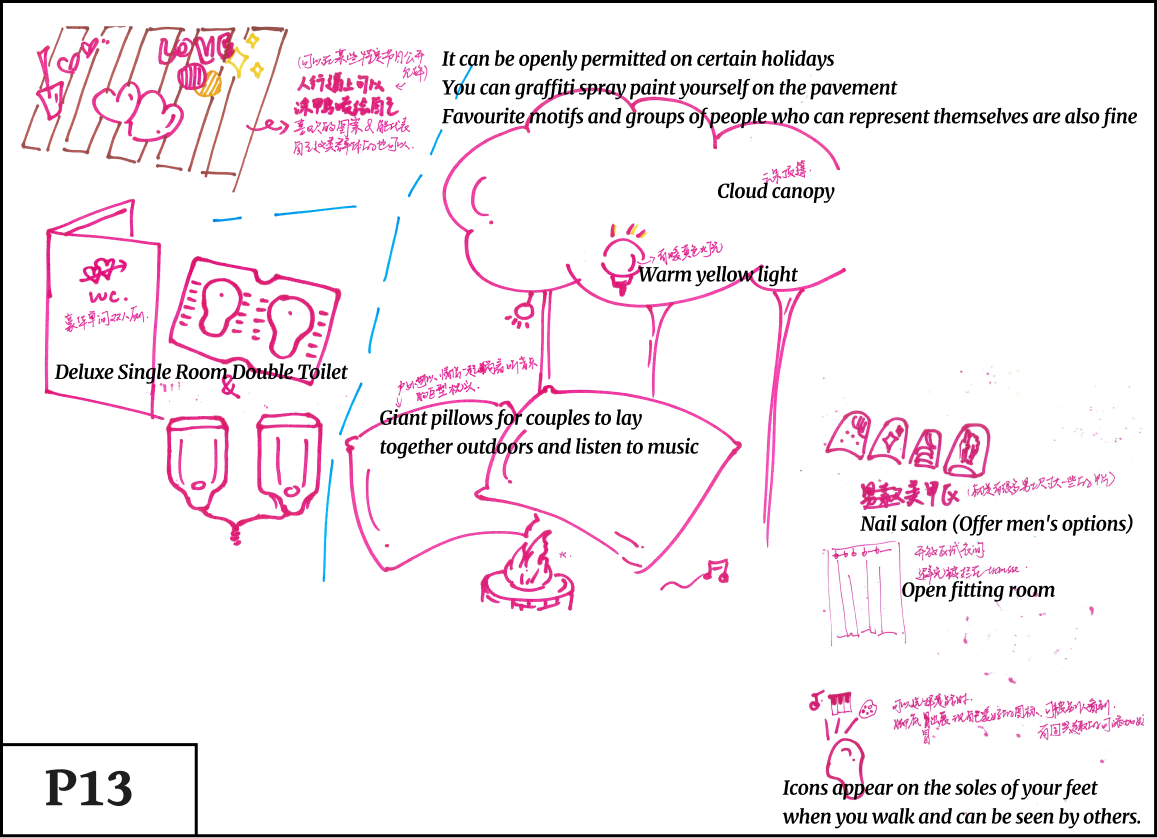}
    \Description{}
    \caption{}
        \label{fig:PP13}
    \end{subfigure}
    \begin{subfigure}[t]{0.49\textwidth}
    \centering
    \includegraphics[height=5.5cm]{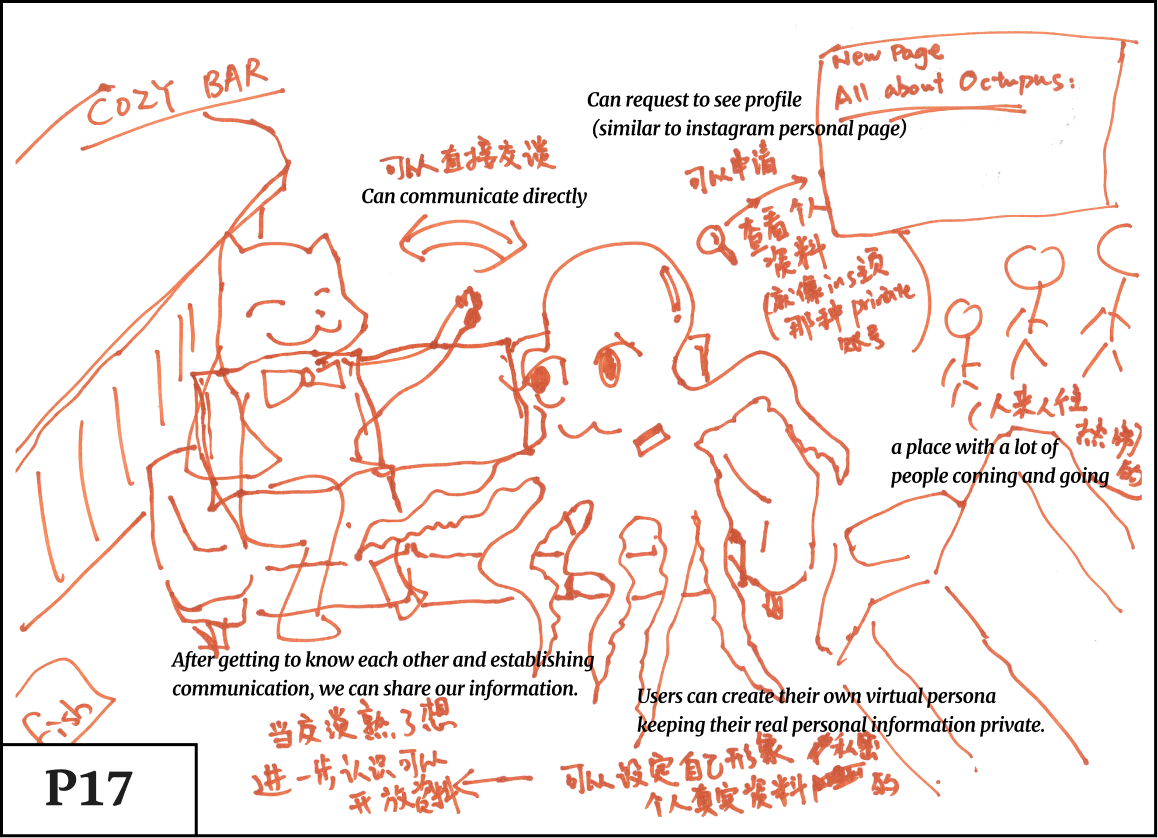}
    \Description{}
    \caption{}
        \label{fig:PP17}
    \end{subfigure}
\caption{Within a small, intimate space: (a) Comfortable and dreamy intimate spaces, eccentric designs, and immersive special effects that seek like-minded people (Workshop 3, P13); (b) A cozy lobby offering chat rooms with various themes that people can choose (Workshop 3, P17).}
\Description{Within a small, intimate space: (a) Comfortable and dreamy intimate spaces, eccentric designs, and immersive special effects that seek like-minded people (Workshop 3, P13); (b) A cozy lobby offering chat rooms with various themes that people can choose (Workshop 3, P17).}
\end{figure}

\textbf{Small Intimate Space.}
Participants emphasized the value of small, intimate virtual spaces in fostering close relationships, particularly for marginalized groups. Such environments, aligned with prior research~\cite{10.1145/3411764.3445729}, were seen as more conducive to emotional expression and cohesion than larger, open settings.
Proposals included \textit{``cozy, dimly lit settings''} designed to encourage vulnerability and genuine communication. P5 envisioned a secure tent where \textit{``close interactions occur naturally without loud voices, enhancing relaxation and social support.''} Other designs incorporated nostalgic elements, such as a \textit{``warm-lit, cloud-canopy space with bean bags''} (P13, Cis Female, Heterosexual; visualized in Figure~\ref{fig:PP13}), or a \textit{``lively hall with plush sofas and a cozy bar''} (P17, Cis Female, Pansexual; visualized in Figure~\ref{fig:PP17}). These intimate settings provide a sense of warmth, safety, and emotional connection, facilitating authentic interpersonal exchanges. 

\subsubsection{Designing for Co-Building Supportive Community.}~\label{sec:coconstruct}
Participants also proposed user-driven collaborative creation approaches in Metaverse spaces to foster community bonds, enhance belonging, and promote shared ownership. 

\begin{figure}[h]
\centering
    \begin{subfigure}[t]{0.49\textwidth}
    \centering
    \includegraphics[height=5.5cm]{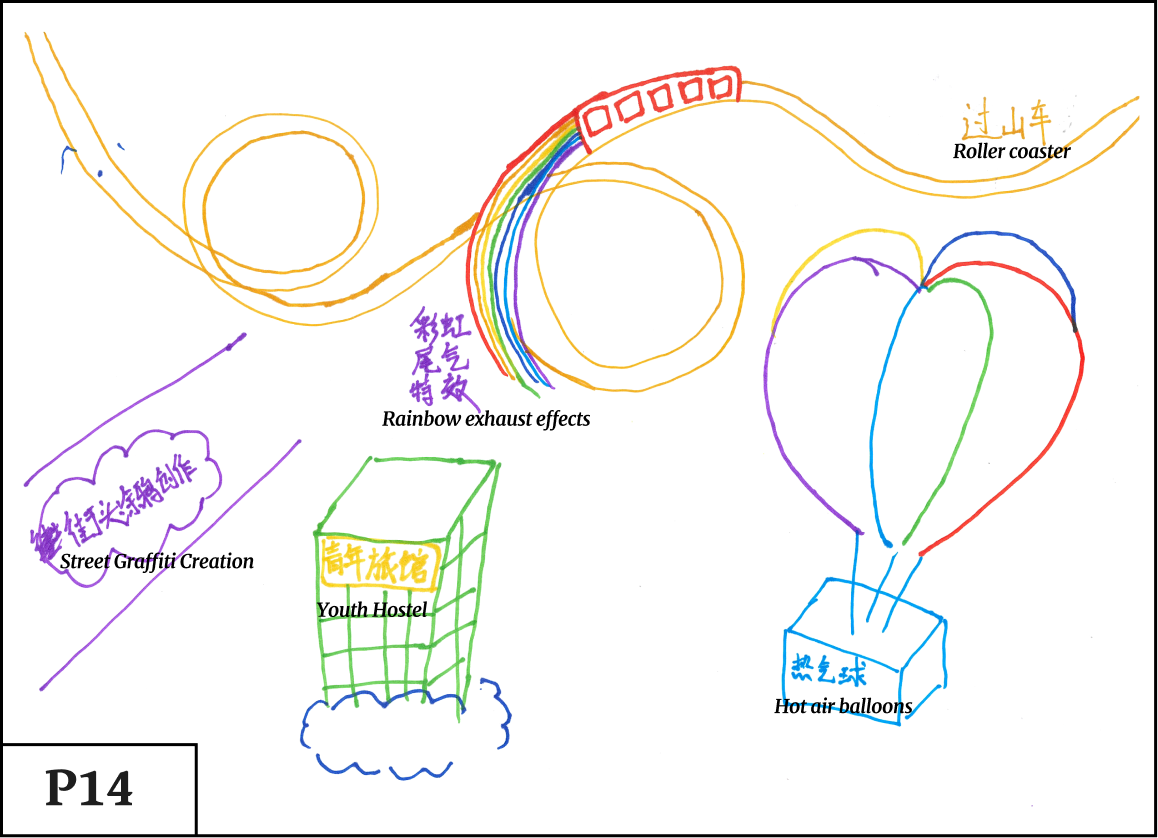}
    \end{subfigure}
\caption{Dreamy transition spaces and gorgeous surrealist style (Workshop 3, P14).}
\Description{Dreamy transition spaces and gorgeous surrealist style (Workshop 3, P14).}
    \label{fig:PP14}
\end{figure}

\textbf{Collaborative Creation System.}  
Many participants envisioned collaborative systems for \textit{``interactive designs, productive tools, and streamlined workflows.''} These systems would enable users to create, and modify, and manage resources such as \textit{``roads, buildings, styles, avatars, and even behavioral data''} (P18, Cis Male, Unsure). 
Participants highlighted one innovative feature in such a system, incorporating nuanced, non-verbal interactions, to enhance user engagement. 
For instance, P14 suggested using gestures with multiple brushes for creative sketching, and \textit{``it supports co-creating public landscapes with others that are difficult to achieve in the physical world, for example, a rainbow waterfall.''} (P14, Cis Female, Bisexual; visualized in Figure~\ref{fig:PP14}). Similarly, P15 (Cis Female, Lesbian) envisioned \textit{``sketching trajectories of body gestures''} to create performance art in which \textit{``dancers’ movements leave colored lines in the air, resembling three-dimensional brush strokes in VR.''} Such innovations integrate physical gestures with virtual tools, transforming artistic performances into immersive, collaborative experiences.
In addition, participants highlighted the need for comprehensive workflows that support diverse behaviors in creative applications. P17 (Cis Female, Pansexual) proposed the creation of \textit{``land art''}—human-made, environmentally interactive art integrated within natural landscapes. She elaborated that \textit{this process should involve exploring extended artistic boundaries, leveraging materials for land art, and fostering collaboration with both creators and audiences.''} These innovative approaches would significantly enrich the shared resources in the Metaverse, enabling users to contribute meaningfully to its content. Through the co-creation process, users could forge deeper connections with the Metaverse, perceiving it as an extension of their physical communities. 
While the proposals lack detailed implementation strategies, they underscore participants' enthusiasm for collaboratively shaping and influencing shared Metaverse spaces. 

\textbf{Meaningful Co-Constructing World.} 
Building upon the concept of collaborative creation, participants envisioned innovative methods for co-constructing virtual worlds that foster collective engagement and shared ownership.
Participants proposed hyper-futuristic methods of \textit{``archiving individual and collective memories''}, using \textit{``volumetric cinema technology by 3D reconstruction''} enabling the preservation of \textit{``monumental moments with permanent time and scenario''} (P4, Cis Male, Gay). This technology could create a distinctive digital landscape where users engage with archived memories through interactive behaviors, such as taking photos with reconstructed scenes displayed in public spaces. 
Participants also emphasized the value of collective memory archives as repositories for shared social connections and cultural continuity. 
For instance, P16 (Cis Male, Heterosexual) envisioned a \textit{``center library''} that would serve as both universal resource access to every inhabitant and a repository of the Metaverse’s collective spirit, reinforcing a sense of shared ownership and identity among virtual inhabitants. 

\begin{figure}[h]
\centering
    \begin{subfigure}[t]{0.49\textwidth}
    \centering
    \includegraphics[height=5.5cm]{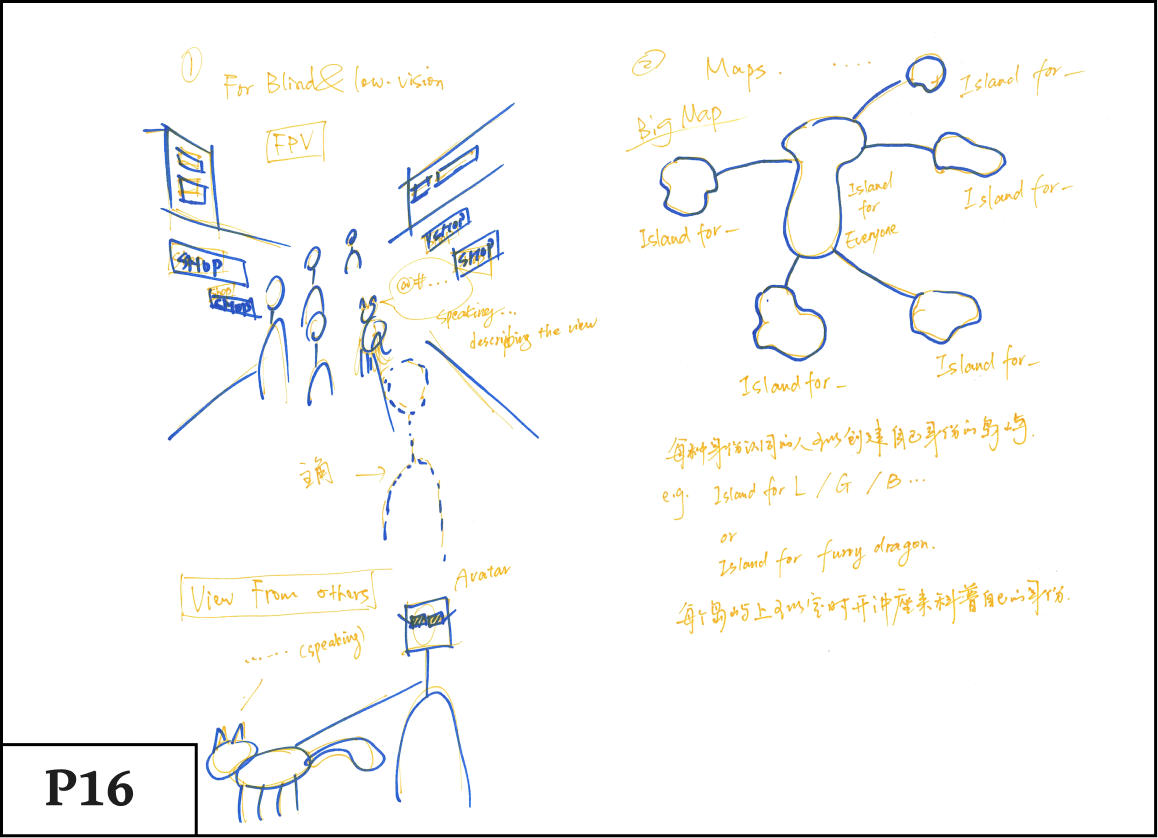}
    \end{subfigure}
\caption{Islands serving as co-creation spaces for different groups (Workshop 3, P16).}
\Description{Islands serving as co-creation spaces for different groups (Workshop 3, P16).}
    \label{fig:PP16}
\end{figure}

In addition to memory archiving, participants envisioned user-driven \textit{``open worlds''}—undeveloped areas within the Metaverse where diverse users collaboratively establish democratic communities. For instance, P18 (Cis Male, Unsure) envisioned \textit{``an uncultivated and unnamed area for co-construction''} (Figure~\ref{fig:PP18}), while P16 (Cis Male, Heterosexual) proposed a self-governed world where \textit{``individuals of different identities could establish their own territories''} (P18, Figure~\ref{fig:PP16}). These concepts prioritize user autonomy and inclusiveness, empowering individuals to shape the Metaverse into a democratic, co-constructed environment. 
However, participants also acknowledged potential challenges in managing shared spaces, particularly the risk of creating \textit{``isolated areas''} that could lead to fragmentation and intergroup conflicts. Effective management strategies would be essential to ensure the development of interconnected, inclusive communities. 
Although ambitious, these visions reflect participants' desire for co-presence and user autonomy within a collaboratively constructed Metaverse.

\subsection{An Unconventional Space with Inclusive Norms}~\label{sec:4.5reformingnorms}
Participants proposed developing spaces that challenge traditional norms, allowing for unconventional interactions that redefine how individuals communicate. These spaces would promote inclusiveness by offering diverse modes of engagement and fostering new ways for LGBTQ+ and cisgender individuals to connect.


     
    \subsubsection{Designing for Empowered Identity Representation.}
    Participants advocated for the development of avatars that reflect a wide range of identities, allowing LGBTQ+ people to express themselves authentically. These inclusive designs allow users to embody their true selves and embrace multiple gender identities and sexual orientations.
    
    
    \textbf{Identity Representation.}
    Consistent with previous studies, participants emphasized the need for diverse avatars and clothing options that transcend gender and identity to enhance communication. P4 (Cis Male, Gay) uniquely extended this idea to include NPCs (non-player characters) in addition to user avatars, using \textit{``characters from traditional myths beyond binary gender''} to challenge views of LGBTQ+ identities as something new. 
    \begin{quote}
        \textit{``A celestial metaverse world could feature figures from traditional mythology, such as Avalokitesvara, who are often portrayed as androgynous. This representation would show that `advanced' views on gender have ancient roots.''} 
    \end{quote}
    By showing the historical depth of gender fluidity and androgyny, this design could encourage wider acceptance and challenge stereotypes. 
    
    Participants also identified the need for \textit{``selective identity management with customization,``} such as an \textit{``interactive information panel''} hovering over avatars. This feature supports selective identity disclosure: \textit{``If one feels safe and comfortable in this environment, I can choose to display my identity to others''} (P18, Cis Male, Unsure). Similarly, participants suggested the platform's responsibility in \textit{``offering an option for either expressing gender identity or skipping this.''} This vision highlights the importance of giving users control over how their identities are displayed according to their situation and context.

\subsubsection{Designing for Unconventional, Democratic Space.} 
Participants highlighted the need for spaces that challenge traditional social norms, enabling new forms of interaction. These spaces would provide LGBTQ+ individuals with a platform for self-expression and connection, free from the constraints of conventional expectations. 

\textbf{Innovative Redesigns for Inclusiveness.}
Participants proposed various design ideas that aim to break heteronormative practices and foster inclusiveness. P3 (Cis Female, Lesbian) and P13 (Cis Female, Heterosexual) similarly envisioned inclusive social spaces like \textit{``nail salons for men''} and \textit{``female-friendly car designs''} to challenge rigid gender binaries: \textit{``The driving systems and interior designs of vehicles in virtual spaces should boldly consider women's needs, even though I have limited choices in real life.''} These designs encourage users to rethink traditional roles and behaviors by introducing such spaces. 

\begin{figure}[h]
\centering
    \begin{subfigure}[t]{0.49\textwidth}
    \centering
    \includegraphics[height=5.5cm]{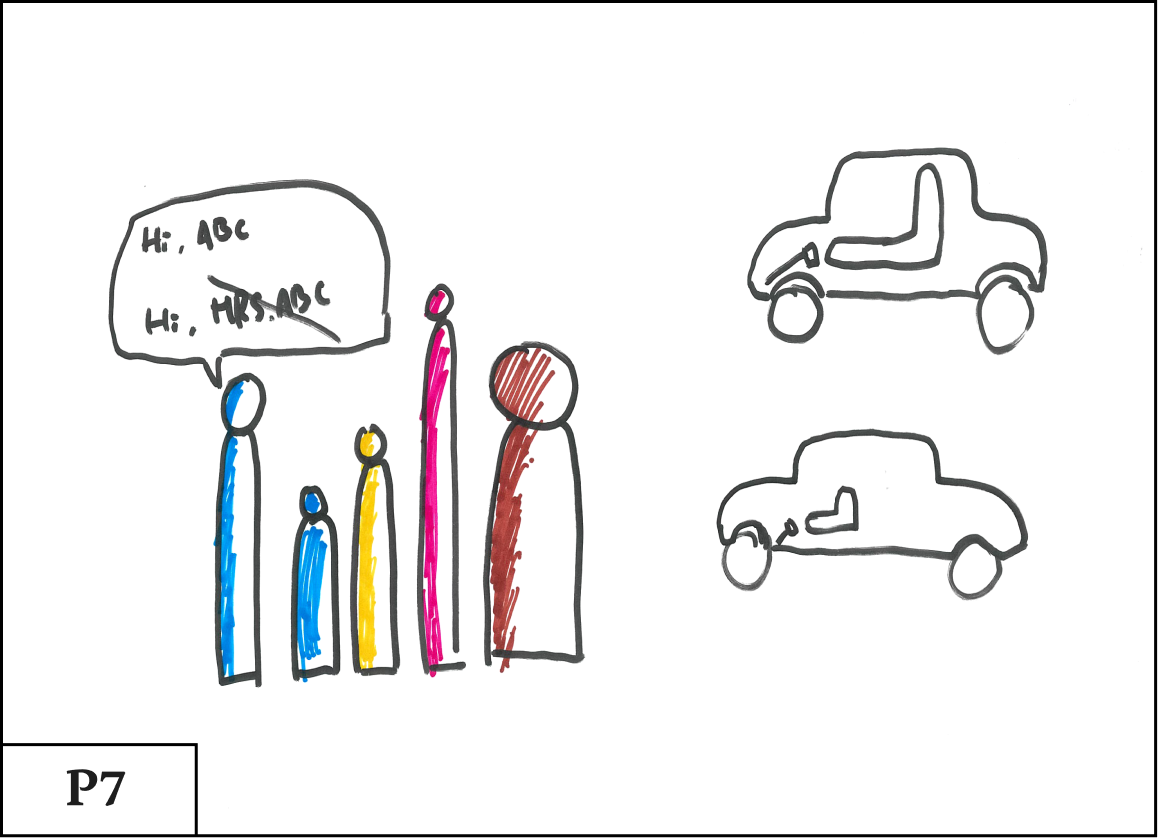}
    \end{subfigure}
\caption{Personalized dress-up experiences instead of LGBTQ+ visibility and design addressing systemic inequality, with a car as an example (Workshop 2, P7).}
\Description{Personalized dress-up experiences instead of LGBTQ+ visibility and design addressing systemic inequality, with a car as an example (Workshop 2, P7).}
    \label{fig:PP7}
\end{figure}

Breaking gender norms in virtual spaces promotes gender equality, enriches personal choices, sparks creativity, and enhances community interaction and awareness. 
Building on the theme of challenging societal norms, P13 (Cis Female, Heterosexual) proposed the \textit{``double toilet room''} concept, normalizing unconventional activities in shared spaces (Figure~\ref{fig:PP13}) by explaining \textit{``When everyone uses the double toilet, such behavior becomes the new norm.''} As she further articulated, 
\textit{``Future metaverse could make us not worry about the challenges posed by biological gender or certain social norms; we can try to break some of the limitations in real life.''} 
This idea underscores the potential of the Metaverse to redefine everyday norms, fostering a sense of belonging for marginalized groups by normalizing diverse behaviors. 
Expanding on the role of practical redesigns, P7 (Cis Male, Gay) highlighted re-designing societal institutions using car design as an example by arguing that \textit{``traditional designs often cater to male users, subtly perpetuating patriarchal norms.''} To counter this, he suggested designing Metaverse cars with \textit{``adjustable features to accommodate diverse body types''} (Figure~\ref{fig:PP7}). 
This practical approach illustrates how inclusiveness can extend to functional designs, respecting individual differences while challenging implicit biases in traditional systems.

\begin{figure}[h]
\centering
    \begin{subfigure}[t]{0.49\textwidth}
    \centering
    \includegraphics[height=5.5cm]{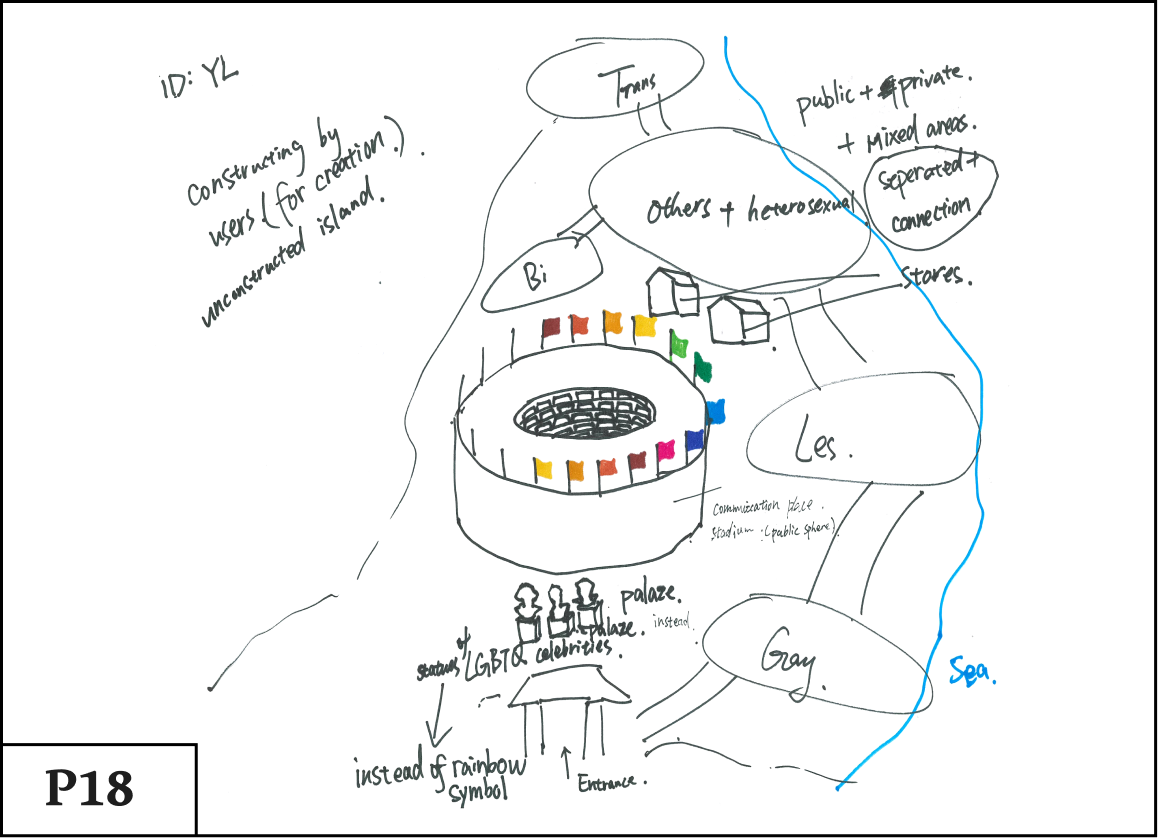}
    \end{subfigure}
\caption{
An inclusive co-creation community incorporating symbolism from historic architectural features (Workshop 3, P18).}
\Description{An inclusive co-creation community incorporating symbolism from historic architectural features (Workshop 3, P18).}
    \label{fig:PP18}
\end{figure}

\textbf{Inclusive Architectural Typology.} 
    Participants also highlighted the symbolic power of architectural forms and spatial organization in fostering inclusiveness and equality. P18 (Cis Male, Unsure) suggested designing a central social space inspired by the Roman Colosseum (Figure~\ref{fig:PP18}), explaining 
    \begin{quote}
        \textit{``Designing an LGBTQ venue in the Metaverse is all about celebrating diversity and inclusiveness. Imagine a space filled with vibrant colors and dynamic virtual environments that capture the energy and variety of the LGBTQ community. We can incorporate symbolic elements like the rainbow flag and different cultural icons to create a safe and welcoming atmosphere[...] This venue wouldn't just host virtual gatherings, art exhibitions, and educational events; it would also encourage interaction and conversation. Participants would feel a real sense of belonging and be free to express themselves in this digital world.''} 
    \end{quote} 
    This concept repurposes a historical symbol of conflict into a platform for equality, self-expression, and shared identity. 
    These proposals reflect the participants’ belief in the transformative potential of Metaverse architecture to symbolize and enact equality.


\section{Discussion}
    This study explores how the design of the future Metaverse can address communication challenges between LGBTQ+ people and cisgender people. 
Despite our study is in mainland China, where a dominant heteronormative culture with largely invisible LGBTQ+ people making it a representation of communication challenges~\cite{2016survey}, these communication challenges prevalent across nations~\cite{2021survey}. Thus, our findings hold universal significance beyond China alone.
By analyzing discussions and visualizations of an envisioned future Metaverse in our workshops, we gained deeper insights into the communication barriers and needs, particularly those experienced by LGBTQ+ communities. 

\hl{
    Although our workshops anticipated hyper-futuristic visions of Metaverse, the proposed solutions were predominantly pragmatic. Most of these solutions were immediate responses to current challenges faced by participants. 
    In the following, we discuss broader implications for LGBTQ+ HCI research, then follow with implications.
    
}

\subsection{Design Suggestions to Enhance Communication Between LGBTQ+ and Cisgender People in Future Metaverse Space}~\label{sec:designsuggestions}
\hl{
    With broad immersive features, virtual spaces could surpass traditional online communities in supporting LGBTQ+ people, through such as social support~\mbox{\cite{acena2021my,freeman2022disturbing,li2023wecried,freeman2022acting,freeman2022re}} (e.g., emotion, information, esteem), and enhanced self-efficacy~\mbox{\cite{schulenberg2023creepy,freeman2022acting,freeman2022re}} (e.g., embodied identity). 
 Comparably, our research identifies new capabilities in addressing communication challenges between LGBTQ+ people and cisgender people. We summarize key new findings below: 
}

\subsubsection{Designing Scenes for Co-Presence Using Nuanced Interaction to Build Collaborative Communication.}~\label{sec:co-shaping} 
A key communication challenge was the lack of meaningful interaction, as the invisible LGBTQ+ communities. 
Little research has been paid to how virtual features can facilitate 
structured interactions between LGBTQ+ and cisgender people—literature suggests that are crucial for prejudice reduction when conducted under optimal conditions~\cite{allport1954nature}. 
Comparably, our finding proposes practical \textbf{interactive features that foster collaborative co-presence that create shared experiences and collective memories}. These designs span multiple interaction 
modes: disseminate knowledge, ideas, and a shared memory (Section~\ref{sec:4.2communication}), collaborative approach encompasses a variety of activities, including creating and modifying resources, interactive performances, participative memory archives, and co-constructing spaces (Section~\mbox{\ref{sec:coconstruct}})—elements that are integral to the Metaverse experience. 
By reveal these, we contributes designing for solidarity within diverse, multi-faceted populations~\mbox{\cite{10.1145/3449162,doi:10.1177/146144804047513}}. 

As DeVito and Walker emphasize, keeping ``the locus of power close to the community''~\mbox{\cite{10.1145/3449162}} is crucial—our participants' designs show how collaborative co-presence can distribute agency across groups while centering LGBTQ+ voices in shaping these shared experiences. We contribute a design approach that shifts focus from 
individual identity expression to collective experience building as a 
foundation for cross-group communication. 
Moreover, by promoting co-presence through collaborative processes  (Section~\ref{sec:4.4co-futurescenes}), our findings offer a new approach to shaping social norms. Rather than accepting existing norms or creating isolated counter-spaces, these designs envision norm transformation through ongoing, structured, collaborative engagement.  
Overall, our findings offer valuable insights into how the Metaverse can evolve beyond just being a medium for interaction; it can serve as a collaborative platform that actively shapes social dynamics and promotes inclusiveness. 

\subsubsection{Designing Rich and Dynamic Proxemics to Ensure a Safe Community Foundation of Communication.}~\label{sec:dynamicproxemics}
Another communication challenge identified by participants is the lack of trust, with many LGBTQ+ participants expressing concerns about being judged or abused in virtual spaces. 
Current studies remain underexplored is how safety mechanisms can be designed to evolve dynamically based on such interpersonal cross-group communication contexts. 
\hl{
To this regard, our findings reveal \textbf{the need for dynamic proxemics—flexible social distances that evolve during interactions to protect privacy and safety}. 
This highly dynamic or flexible visualization of social distance can be represented through various elements, such as avatar color changes (Section~\mbox{\ref{sec:flexibleinteraction}}), social networks (i.e., contact by invitation), manageable boundaries (i.e., social distance with others), spatial locations (Section~\mbox{\ref{sec:socialfeatures}}). 
These elements can evolve in ongoing social interactions, acting as indicators of trust, privacy, and safety within the Metaverse space. 
Compared to current designs on static mechanisms (e.g., fixed protective shields or global trust systems with predefined settings~\mbox{\cite{zheng2023understanding,10.1145/3359202,chen_people_2024}}), our findings propose a \textbf{context-aware approach grounded in cross-group communication}. 
Specifically, the proposed dynamic proxemics were envisioned to leverage nuanced social cues from non-verbal interactions (e.g., body language, gestures, emotions) to enhance interpersonal dynamics. 
}

Aligning with works on social trajectories and interpersonal distances, we extend prior studies (e.g.,~\mbox{\cite{10.1145/3411764.3445729,10.3389/fpsyt.2020.00561,li2023wecried}}) 
By revealing how proximity shapes what is considered safe and acceptable in virtual environments, our findings extend prior work on embodied safety~\mbox{\cite{freeman2022re,freeman2022acting}} and selective visibility~\mbox{\cite{yao2023beyond}}. We demonstrate that safety in cross-group communication is not a binary state but a continuous negotiation mediated by spatial relationships. This contributes a nuanced understanding of how Metaverse design can support progressive trust-building while protecting vulnerable users. 

\subsection{Implications for Inclusive Future Design under Power Geometries}~\label{sec:powergeometry}

    Our participants expressed a strong desire to take control of space ownership (Section~\ref{sec:4.4co-futurescenes}) and construction to challenge and redefine conventional heteronormative design norms (Section~\ref{sec:4.5reformingnorms}). For instance, a ``male nail salon,'' and spatial typologies like ``circular walkways or stages,'' ensure equal spatial positioning for all avatars and promote participation. 

    To better understand this need for control over space, we refer to Doreen Massey's concept of ``power geometries.'' Massey's framework explains how spatial mobility and access are influenced by power structures, revealing persistent power inequalities in the context of modernity and globalization~\cite{relph1976place}. Dourish extended this theory into digital spaces, illustrating how power imbalances are recreated in online environments where spaces are designed or formed~\cite{dourish2006re}. Similarly, Scheuerman et al. highlighted that marginalized groups face the reproduction of these power dynamics from physical to virtual spaces, particularly within online communities~\cite{scheuerman2018safe}. Our findings that emphasize the desire to reclaim ownership of space align with previous studies on power geometries, showing that marginalized groups seek to resist and reshape dominant social conventions and institutional structures~\cite{mckenna2020resistance}. 
    Prior studies have applied the concept of power geometries to discuss the challenges faced by transgender and LGBTQ+ people in virtual spaces~\cite{scheuerman2018safe}. Our research expands on this by exploring power dynamics in the design and construction of Metaverse spaces, especially between LGBTQ+ and cisgender people. Through the framework of power geometries, we raise two critical perspectives: 

\textit{\textbf{Using Combined Perspectives to Unveil Power Dynamics and Empower Marginalized Voices.}} 
Designing ownership systems in virtual spaces allows marginalized groups to reclaim control over their technological environments, addressing inherent privileges embedded within design norms~\cite{yao2023beyond}. Hall et al. argued that to fully understand inequality, the privileges of the dominant group must be scrutinized as much as the oppression faced by others~\cite{hall2020other}. Similarly, HCI researchers have advocated for design processes that prioritize the needs of marginalized communities, giving them greater control over how virtual spaces are constructed and experienced~\cite{scheuerman2018safe}. 

Our study shows that integrating perspectives from LGBTQ+ and cisgender groups is essential for fostering effective communication and collaboration in such future workshops. By creating additional channels for dialogue, we can compare perspectives from LGBTQ+ communities with cisgender groups to identify differing opinions (i.e., collaboratively identifying challenges and envisioning during discussion). These rich differences can illuminate how various power dynamics shape participants' ideas between different groups. 
We argue that introducing this joint-group perspective in envisioning future scenarios is crucial for decentralizing authority, that is, exemplifying the \textit{``refocus the marginalized population as the authority''}~\cite{harrington2019deconstructing}. The actions using this approach could benefit both LGBTQ+ and cisgender people, offering more balanced and practical design guidance for future dialogues and collaboration. 

\textbf{\textit{Considering the Communities' Context and Understanding Individuals as Root of Future Design Methods.}} 
\hl{Although the future envisioning of design workshops has sometimes been dismissed as ``impractical'' to reflect the needs of marginalized communities~\mbox{\cite{harrington2019deconstructing}}, Hardy et al. argue that such critiques stem from the limitations of user-centric models rooted in capitalist industries, which are detached from the deep pragmatic foundation of future workshops embedded in marginalized communities~\mbox{\cite{hardy2022lgbtqfuture}}. 

Our study also anchors this vision by revealing the power dynamics of spatial construction within the privileged framework of participatory design, highlighting the rich individual differences within communities. 
From our design sessions, we noticed how proposed designs often normalized oppression, such as debates over identity control. However, these views were challenged by participants' first-hand experiences, bridging communication gaps and creating a political space where, as Butler posits, ``the body is politics''~\mbox{\cite{butler2011bodies}}. By envisioning an ideal future beyond heteronormative designs, the future workshop fosters safe environments and authentic dialogues that support marginalized groups while critiquing power inequalities. 
As Harrington et al. emphasize the importance of historical context and personal experience~\mbox{\cite{harrington2019deconstructing}}, we advocate for reclaiming the foundational principles of future workshops, with a focus on diversified individuals distinguished through critical, authentic communication rooted in pragmatism. }

\section{Limitation and Future Work}
\hl{
Our work derives thematic topics from sketches and discussion through collective conversation. Given the qualitative nature of our research, our findings are not intended to generalize across a broad population (e.g., the wider LGBTQ+ identity spectrum). Based on information shared by participants, the sample also tends to be well-educated and financially stable.
Nevertheless, our qualitative approach provides in-depth insights into the communication needs of LGBTQ+ individuals and offers a critical perspective on future Metaverse-driven inclusiveness, aiming to inspire further research. Future studies could extend our rich findings by evaluating the effectiveness of diverse analytic designs through user studies and incorporating a wider range of LGBTQ+ identities and stakeholders. 

Moreover, our study addresses the substantial relevance of communication challenges for LGBTQ+ individuals within the HCI community. 
However, the focus on ``application areas'' for LGBTQ+ insights—rather than positioning LGBTQ+ individuals as a central ``research focus''—aligns with critiques in queer HCI, which argue against limiting LGBTQ+ issues to mere utility in application spaces~\mbox{\cite{10.1145/3613904.3642494}}. 
Nonetheless, our findings framework centers on LGBTQ+ perspectives, emphasizing the diverse voices of transgender and non-binary participants to cultivate an environment of mutual respect and understanding. By drawing on ``radical design'' methodologies~\mbox{\cite{lindley2015back}}, this approach explores alternative social contexts and idealized social conditions as meaningful, actionable implications, moving beyond purely utilitarian perspectives. 

Based on the findings of this study, future work can incorporate the design ideas proposed by our participants and conduct user studies to assess the applicability and effectiveness of these suggestions. Many of the design recommendations, such as spatial scenes and some of the dynamic social features, can be easily implemented. Future studies may also adopt more realistic daily communication scenarios to investigate the challenges faced by LGBTQ+ groups. These studies could explore how to recreate these scenarios in the Metaverse using some of the design suggestions provided in our findings to lower communication barriers.
}


\section{Conclusion}
    As virtual technologies continue to gain popularity, designing a Metaverse space that aims for equality and inclusiveness is becoming increasingly essential. In this study, we invited participants from both LGBTQ+ and cisgender people to collaboratively envision the future inclusive Metaverse space where the communication between LGBTQ+ and cisgender people can be more effective. The solutions proposed by participants are based on mutual understanding, developed through a full discussion to address the current communication challenges between LGBTQ+ and cisgender people. Subsequently, we collected opinions from imaginative ideas and creative sketches to form thematic solutions. 

In our work, participatory design offers a valuable channel for linking extensive voices and imaginations across LGBTQ+ and cisgender communities. Meanwhile, future envisioning of design workshops provides a powerful approach to questioning current technology forms that reproduce power inequality. 
Our findings offer innovative insights into power geometries among LGBTQ+ and cisgender people and provide novel design suggestions for Metaverse spaces. This study aims to empower LGBTQ+ groups by promoting greater equality and equal rights in the future design process for the future inclusive Metaverse space.

\section{Acknowledgment about the Use of LLM}
The authors would like to acknowledge the use of the generative AI tool in this work. Specifically, \textit{GPT-4o mini} by OpenAI was utilized to assist in language refinement, including grammar and style corrections of existing manuscript text.





\bibliographystyle{ACM-Reference-Format}
\bibliography{dreamthedream}



\end{document}